\pdfoutput=1
\documentclass[11pt]{article} 

\usepackage[utf8]{inputenc} 

\usepackage{geometry} 
\usepackage{graphicx} 
\usepackage{graphics}

\usepackage{tikz}

\tikzset{
    vertex/.style = {
        circle,
        fill            = black,
        outer sep = 2pt,
        inner sep = 1pt,
    }
}

\usetikzlibrary{matrix,chains,positioning,decorations.pathreplacing,arrows}

\usepackage{array} 

\usepackage{enumerate}

\usepackage{amsmath}
\usepackage{amssymb}
\title{Introduction to Solving Quant Finance Problems with Time-Stepped FBSDE
and Deep Learning}
\author{Bernhard Hientzsch\thanks{Corporate Model Risk, Wells Fargo Bank,
bernhard.hientzsch@wellsfargo.com}}

\begin{document}
\date{11/25/2019}
\maketitle
\setcounter{secnumdepth}{5}
\begin{abstract}

In this introductory paper, we discuss how quantitative finance problems under
some common risk factor dynamics for some common instruments and approaches can
be formulated as time-continuous or time-discrete forward-backward stochastic
differential equations (FBSDE) final-value or control problems, how these final
value problems can be turned into control problems, how time-continuous
problems can be turned into time-discrete problems, and how the forward and
backward stochastic differential equations (SDE) can be time-stepped. We obtain
both forward and backward time-stepped time-discrete stochastic control problems
(where forward and backward indicate in which direction the $Y$ SDE is
time-stepped) that we will solve with optimization approaches using deep neural
networks for the controls and stochastic gradient and other deep learning
methods for the actual optimization/learning. We close with examples for the
forward and backward methods for an European option pricing problem. Several
methods and approaches are new.
\end{abstract}

\section{Introduction}

This paper is structured as follows: in section 2, we quickly discuss the
general risk factor dynamics used. In section 3, we describe the prototypical
instruments and instrument features treated: Europeans, Barriers, and
Bermudans/Exercise opportunities.
Section 4 states what we are interested in computing. Section 5 shows how one
can obtain (continuous time) FBSDE formulations and FBSDE final value problems
from a variety of sources and approaches. Section 6 shows how one can obtain
(continuous time) FBSDE stochastic control formulations. Section 7 shows how one
can obtain (discrete time) FBSDE stochastic control formulations for the
introduced financial instruments. Section 8 describes how these \mbox{discrete}
time FBSDE stochastic control problems can be solved by deep neural networks and
deep learning.  Section 9 presents the application of the forward and the
backward methods to European option pricing problem for a one-dimensional
example which allows good visualization and understanding. Section 10 concludes.

\section{Risk Factor Dynamics}

The vector of risk factors $X_{t}$ under consideration is assumed to follow the
SDE:
\begin{equation}
dX_{t} = \mu\left( t,X_{t} \right)dt + \sigma_N\left(t,X_{t}\right)dW_{t}
\end{equation}

The operator connected to the risk factor dynamics is:
\begin{equation}
{\cal L}_t u\left( t,x \right) :=  
\frac{1}{2}\text{Tr}\left(\sigma_N\sigma_N^{T}\left( t,x \right)\left(
	\text{Hess}_{x}u \right)\left( t,x\right) \right) 
+ \mu\left( t,x \right)\nabla u\left( t,x \right)
\end{equation}
where $\text{Hess}_{x}u$ is the Hessian matrix. 

For the case of constant volatility, constant interest rate, risk neutral
Black-Scholes on a single underlier, $\mu(t,X_t)=r X_t$ and
$\sigma_N(t,X_{t})=\sigma_{BS} X_{t}$, and ${\cal L}_t u = \frac{1}{2}
\sigma_{BS}^2 X^2 u_{xx} + r X u_x$.

Most often, we will assume that the dynamics is given under the risk neutral
measure where $\mu(t,X_t)= r(t) X_t$ for tradeable components of $X_t$ or under
numeraire measures in which $\mu(t,X_t)$ is zero and the components of $X$ are
measured in units of numeraire.  We also typically assume (in particular for
the trading strategy set-up) that the volatility is given in log-normal terms
$\sigma_N(t,X_{t})=\sigma_{LN}(t,X_{t}) X_t$ (with multiplication understood
elementwise).
 
Typically one assumes at least one locally riskless basic security (money market
instrument or bond) with an equation like the following:
\[dB = r\left( t \right)B\left( t \right)\text{dt}\]
or alternatively a risk-free zero-coupon bond or similar instrument (the
expression and/or SDE for such bonds depends on the chosen model).

\section{Instruments covered}

In this introduction, we will only cover relatively simple instruments with a
few key features. Many instruments with more complicated features can be treated
very similarly with the same approaches and ideas.

\subsection{Europeans}

At a given payment time $T^P$, called typically instrument maturity, a European
option instrument pays $g(X(T^P))$.
$X$ is the vector of the risk factors.  (It might well be that the final payoff
does not depend on any risk factor or on only one.) For a stock-option, $X$
would be the stock price.
For a short-rate model, the risk factor could be the short rate $r(t)$, or if
the formulation requires a tradeable financial instrument, it could a bank
account $B(t)$ or zero coupon bond $P(t)$ under that short rate model, and the
quantity that is used to determine the payment amount could be a forward rate
(which can also be written in terms of a ratio of two bond prices) $L$.

The two most common basic payoffs are call and put options on some single
underlier $X_i(T^P)$ with strike $K$, paying $\mathsf{max}(X_i(T^P)-K,0)$ and
$\mathsf{max}(K-X_i(T^P),0)$ respectively. 

\subsection{Simple Barriers}

Barrier options are options that will pay only if some barrier level or region
is touched (knocked-in) or not touched (knock-out), or change final payout upon
touching a barrier. The simplest barrier options are those with a
single barrier at some constant level which is active for the
entire life of the instrument.

Once again, two standard examples: A standard up-and-out barrier call option
with upper barrier $B$ will pay the final call payoff unless the underlier $S$
of the option was observed at a level $S\geq B$ during the life of the option
and otherwise will pay nothing. A standard up-and-in barrier call option
with upper barrier $B$ will pay the final call payoff only if the underlier $S$
of the option was observed at a level $S\geq B$ during the life of the option
and otherwise will pay nothing. 

In terms of simulation or simulation-like approaches, one follows the risk
factor simulation until maturity or barrier breach (whatever comes first) and
then uses the final value or the value of the knocked-in instrument on/in the
barrier (or zero, if there is no knock-in).

\subsubsection{Special case: Treatment as European}

We will explain the case of an upper barrier at a constant level $B$:
Call $\textrm{PBreach}(X_T;X_0)$ the probability that the barrier was breached
given initial and final value $P(X_t\geq B|0<t<T,X_0,X_T))$ and call the payout when
triggered $g_B(X_T)$ and when not triggered $g_{NB}(X_T)$. If the final value of
$X$ is not beyond the barrier, the final value of the instrument is either
$g_B(X_T)$ with probability $\textrm{PBreach}(X_T;X_0)$ or $g_{NB}(X_T)$ with
probability $(1-\textrm{PBreach}(X_T;X_0))$; if it is beyond the barrier, it
will be $g_{B}(X_T)$.\footnote{One can similarly treat lower and double barriers.
Barriers with nonconstant level can be treated if the
appropriate breach probabilities under that particular model can be computed
explicitly accurately and efficiently enough which is typically only possible
in special cases.}

For purposes of valuation as of time 0, the value of the barrier option 
will agree with the value of an European option with the final value
\begin{equation}  
g(X_T) = 
\left\{
\begin{array}{lr} 
g_{NB}(X_T) (1-\textrm{PBreach}(X_T;X_0)) + g_B(X_T)
\textrm{PBreach}(X_T;X_0) & \textrm{if not  } X_T \geq B \\
g_B(X_T) & \textrm{if  } X_T \geq B \\
\end{array}
\right.
\end{equation}

This can be solved just like any other European option pricing problem. 
Notice that this will only give the correct price if the barrier has not been
breached at valuation time. 

Yu, Xing, and Sudjianto \cite{yu2019deep} have used a variant of this approach
to solve some barrier options with the standard European deepBSDE approach. 
 
\subsection{Exercise opportunity}

Assume that at some time $t_E<T$, the holder of the instrument has a choice to
either exercise into an immediate payment worth $g_E(X_{t_E})$ (or into an
instrument expected to be worth $g_E(X_{t_E})$) or continuing to hold on to the
instrument with the given final payment $g(X(T^P))$.

Exercise opportunities are often handled by either computing an expected holding
value/continuation value $\mathsf{hv}(X_{t_E})$ and exercising when the exercise
value is larger than the holding value ($g_E(X_{t_E})>\mathsf{hv}(X_{t_E}) $)
or by defining an exercise strategy $\mathsf{es}(X_{t_E})$ that is only true/one
when the instrument should be exercised and false/zero otherwise. Given some
holding value function or exercise strategy function, an exercise opportunity can be
directly simulated and pricing happens like the barrier case (where different
actions are taken depending on whether the barrier was hit or not).

Proceeding from the last exercise opportunity to the first, the case of finitely
many exercise opportunities can be reduced to the case of a single exercise
opportunity. Exercise time intervals can be approximated by appropriately
frequent discrete exercise times. 

\section{Analytics to be computed}

At a minimum, we want to compute the value at initial time with given fixed risk
factor values. (This corresponds to the dynamics being started at $X_0=x_0$
with $x_0$ being those fixed risk factor values.) In many situations, we would
like to compute the value at initial time with risk factor values within a
certain range around some given fixed values (for sensitivities and other
purposes). This can be achieved by modeling $X_0$ as a random variable with the
appropriate domain, for instance. 

The methods that we will present will compute simulated instrument values along
simulated paths. Forward methods will give us simulated instrument values
conditional on the shared past. Backward methods will give simulated instrument
values that take future values of the risk factors and of the instrument value
into account.  To convert to instrument values conditional on the shared past,
an adapted projection or approximation needs to be computed from the simulation
results of the backward methods. 

In general, it would be useful to determine the instrument value at certain
intermediate times over a certain range of risk factors. This can be achieved for instance by
starting the computation at a future time $t_0$ with random $X_{t_0}$ with the
appropriate domain of interest, but potentially also with other approaches. For
instance, this can be used to compute holding values for exercise opportunities.

\section{Obtaining time-continuous FBSDE final value problems for Quantitative
Finance problems}

A time continuous FBSDE problem has the following form: 

The forward SDE (FSDE) for the dynamics:
\begin{equation}
dX_{t} = \mu\left( t,X_{t} \right)dt + \sigma_N\left(t,X_{t}\right)dW_{t}
\end{equation}

and the backward SDE (BSDE) for the value in terms of volatility scaled values
$Z_{t}$:
\begin{equation}
- dY_{t} = f_Z\left( t,X_{t},Y_{t},Z_{t} \right)dt\  - Z_{t}^{T}dW_{t}
\end{equation}
or in terms of values $\Pi_{t}$:\footnote{If $\Pi_{t}$ measures the units of
securities (rather than the value invested in such) in the portfolio, it would
be $\sigma_{N}$ rather than $\sigma_{LN}$ in the stochastic term of the $Y$ BSDE.}
\begin{equation}
- dY_{t} = f\left( t,X_{t},Y_{t},\Pi_{t} \right)dt\  - \Pi_{t}^{T}
\sigma_{LN}\left(t,X_{t}\right) dW_{t}
\end{equation}

A final value problem adds the final value condition for $Y$
\begin{equation}
Y_{T} = g(X_{T})
\end{equation}

For fixed initial risk factor values, the forward dynamics is completed by 
$X_0=x_0$. For the case of random $X_0$, $X_0$ will be a random variable. 

For short introductions into FBSDE in finance and otherwise, see, for
instance, El Karoui, Peng, and Quenez \cite{el1997backward} and Perkowski
\cite{perkowski2010}.

\subsection{Linear PDE}

A linear partial differential equation (PDE) of the form:
\begin{equation}
u_{t}\left( t,x \right) + {\cal L}_t u\left( t,x \right) - V\left(t,x\right) u\left(t,x\right)
+h\left(t,x\right)= 0
\end{equation}
can be written as a FBSDE with the generator function $f$ as follows: 
\begin{equation}
f(t,X,Y,\Pi) = - V(t,X) Y + h(t,X)
\end{equation}
as a special case of a nonlinear Feynman-Kac theorem (which will be presented
below).

\subsection{Risk Neutral Expectations}

Assume the solution is characterized by 
\begin{equation}
u\left( t,x \right) = {B\left( t \right)E}\left(\frac{ g\left( X_{T}
\right)}{B(T)} \middle| X_{t} = x \right)
\end{equation}
with 
\begin{equation}
\frac{B(t)}{B(T)}=\exp \left( - \int_{t}^{T} r\left( s \right) ds \right)
\end{equation} 

Then the solution $u$ can be written as $u(t,X)=Y_t$ where $X_t=X$ and $Y$
solves a FBSDE with generator function $f$ as follows:
\begin{equation}
f(t,X,Y,\Pi) = - r(t) Y
\end{equation}
 
\subsection{Expectations under Numeraire Measures}

Assume the solution is characterized by
\begin{equation}
\frac{u\left( t,x \right)}{N(t)} = E\left(\frac{ g\left(
X_{T} \right)}{N(T)} \middle| X_{t} = x \right)
\end{equation}
with some deterministic or stochastic dynamics for the numeraire $N(t)$, where
$E$ is in the measure corresponding to the numeraire $N(t)$. Under that measure,
the relative value of the instrument as measured in units of numeraire is a
martingale.
Therefore, the generator function $f$ will be zero. 

The solution $u(t,X)$ will be given as $u(t,X)=N(t) Y_t$ where 
$X_t=X$ and $Y$ solves a FBSDE with a zero generator function. 

\subsection{Nonlinear PDE}

One of the nonlinear Feynman-Kac theorems states (El Karoui, Peng, and Quenez
\cite{el1997backward} and Perkowski \cite{perkowski2010}) that under appropriate
assumptions, the solution of
\begin{equation}
u_{t}\left( t,x \right) + {\cal L}_t u\left( t,x \right) + f\left( t,x,u\left( t,x \right),\nabla
 u\left( t,x \right) \right) = 0 
\end{equation}
is given as the $Y$ solution $Y_t$ of a BSDE
\begin{equation}
- dY_{t} = f\left( t,X_{t},Y_{t},\Pi_{t} \right)dt\  - \Pi_{t}^{T}
\sigma\left(t,X_{t}\right) dW_{t}
\end{equation}

The $\Pi$ solution of the BSDE will be $\Pi_t=\nabla_X u(t,X_t)$.

\subsection{Self-Financing Conditions}

Instead of deriving the BSDE from other formulations, one can directly derive
BSDE from self-financing conditions if one assumes that the components
of $X$ are basic underlying securities (or, at least, that there are enough
instruments that depend on such components of $X$ and that hedge ratios for such
instruments can be computed from the sensitivities to the components).
The corresponding component of $\Pi_t$ (the trading strategy portfolio)
describes how much of the portfolio value is invested in that component of
$X_t$. The remainder of $Y_t$, $\pi_0(t)=Y_t- \Pi \cdot 1$, is assumed to be
invested in cash.

Under the assumptions of risk neutral pricing (among them, that both positive
cash balances (which can be lent out) and negative cash balances (which
corresponds to amounts borrowed) attract the same interest rate $r(t)$ - deterministic or
stochastic), one obtains a BSDE for the portfolio value $Y_t$ with generator
function $f$ as follows:
\begin{equation}
f(t,X,Y,\Pi) = - r(t) Y
\end{equation}

Assuming that positive cash amounts attract an interest rate $r^l(t)$ and
negative cash amounts attract an interest rate $r^b(t)>r^l(t)$ and that 
the components of $X_t$ have a drift of $r^l(t) X_t$ (this setting is called
``differential rates''), then the $f$ generator function in the BSDE is as
follows:
\begin{equation}
f(t,X(T),Y(t),\Pi_{t}) = -r^l(t) Y(t) + (r^b(t)-r^l(t)) \left(\sum_{i=1}^n \pi_i(t) -
  Y(t) \right)^{+}
\end{equation}

Transaction cost can be included if the generator $f$ is allowed to depend on
the time-derivative $\dot{\pi}_i(t)$ of $\pi_i(t)$ as well, with $G(x)=\lambda
|x|^q/q$ with $q \in (1,2]$ and additional terms $-G(\dot{\pi}_i(t))$ for all
the components with transaction costs added to the $f$ generator function in the
BSDE, as in Gonon, Muhle-Karbe, and Shi \cite{gonon2019asset}.
(Alternatively, the transaction costs can be included in the running cost in the
stochastic control problem introduced below.) However, it is easier to handle
transaction costs in the time-discrete setting.

\section{Obtaining time-continuous FBSDE stochastic control problems}

A general stochastic control problem based on some underlying stochastic
evolving factors $\tilde{X}$ asks for a control function $c_t$ that minimizes or
maximizes the functional 
\begin{equation}
\tilde{J}(c) = E\left(\int_0^T \tilde{\mathsf{rc}}(s,\tilde{X}_s,c_s) ds +
\tilde{\mathsf{fc}}(\tilde{X}_T)\right)
\end{equation}
where $\tilde{X}$ follows: 
\begin{equation}
d\tilde{X}_{t} = \tilde{\mu}\left( t,\tilde{X}_{t};c_t \right)dt +
\tilde{\sigma}\left(t,\tilde{X}_{t};c_t\right)dW_{t}
\end{equation}
and $\mathsf{rc}$ and $\mathsf{fc}$ are running cost and final cost functions,
respectively. 
In the standard exposition, $\tilde{X}_0$ is typically given as fixed value, but
part of it could be part of the control and/or a function of
some other components of $\tilde{X}$ where the particular function is
determined as part of the solution of the control problem.

One can also derive a BSDE that characterizes the optimal control for this
stochastic control problem. However, here $\tilde{X}$ will be the concatenation
of the $X_t$ vector and the $Y_t$ value and thus the underlying stochastic
evolving factors will obey
\begin{equation}
dX_{t} = \mu\left( t,X_{t} \right)dt + \sigma\left(t,X_{t}\right)dW_{t}
\end{equation}
\begin{equation}
 dY_{t} = - f\left( t,X_{t},Y_{t},\Pi_{t} \right)dt\  + \Pi_{t}^{T}
\sigma\left(t,X_{t}\right) dW_{t}
\end{equation}
where $\Pi_{t}$ plays the role of a control and the functional to be minimized
or maximized is 
\begin{equation}
J^F(\Pi_t,\Pi^{\mathsf{final}},\ldots) = E\left(\int_0^T
{\mathsf{rc}}(s,{X}_s,Y_s,\Pi_s) ds + {\mathsf{fc}}({X}_T,Y_T,\Pi^{\mathsf{final}})\right)
\end{equation}
 
Any forward FBSDE stochastic control problem of that form occuring in the
literature or in applications can be handled in our approach. 

For a very short introduction into stochastic control problems, see Perkowski
\cite{perkowski2010}.

\subsection{From Final Value problems - Forward Approach}

If the $Y_t$ BSDE is treated in a forward manner, the initial value of $Y$,
$Y_0$ is part of the control (if $X_0$ is a fixed value), while if $X_0$ is
random, $Y_0=Y^{\mathsf{init}}(X_0))$ is a function to be determined as part of
the stochastic control problem. Transaction costs and similar could be treated
as part of the BSDE or as part of the running cost. The final cost will be some
(risk) measure on how well the final value is replicated, for instance the $L_2$
distance
\begin{equation}
\mathsf{fc}(X_T,Y_T) = || Y_T - g(X_T) ||^2 
\end{equation}
but other appropriate risk measures are possible also.

The forward approach for fixed $X_0$ in the time-discrete case (together with
approximating the control with deep neural networks (DNN) and solving the
control problem with deep learning (DL)) was introduced and used in applications
by E, Han, and Jentzen \cite{weinan2017deep} and called ``deepBSDE'' method, a
generic name that we will also use.

To the best of our knowledge, Han, Jentzen, and E \cite{han2018solving} mention
the forward approach for random $X_0$ as a possibility on page 8509 but we are
not aware of any implementation of this method besides our own.

\subsection{From Final Value problems - Backward Approach}

Assume that there would be a way to treat the $Y_t$ BSDE backward in the
time-continuous setting, then one could start with $Y_T=g(X_T)$ and 
evolve $Y_t$ backward until one reaches time 0 or another chosen initial time
$t_0$.

Under those circumstances, the functional to be minimized or maximized would be 
\begin{equation}
J^B(\Pi_t,\Pi^{\mathsf{initial}},\ldots) = E\left(\int_0^T
{\mathsf{rc}}(s,{X}_s,Y_s,\Pi_s) ds +
{\mathsf{ic}}({X}_0,Y_0,\Pi^{\mathsf{initial}})\right)
\end{equation}

If there is a unique solution $Y_t$ to the final value problem, that solution
should be found as minimizer of a functional where the running cost is zero and
the initial cost is given as the variance of $Y_0$ (or any other good measure
of range or size of domain of $Y_0$) if $X_0$ is fixed or the
$L_2$ distance of the $Y_0$ from a to-be-determined function
$Y^{\mathsf{init}}(X_0)$ if $X_0$ is random. 
Alternatively, the
initial cost could be combination of a variance/range measure and risk measure or one could solve a
multi-objective problem with both of those measures as separate objectives. 

The backward method for fixed $X_0$ in the time-discrete case (together with
approximating the control with DNN and solving the control problem with deep
learning), to the best of our knowledge, was introduced and used first in Wang
et al \cite{wang2018deep}.
The backward method for random $X_0$ is new.

\section{Obtaining time-discrete FBSDE stochastic control problems}

\subsection{Time-discretizing time-continuous FBSDE}

Applying a simple Euler-Maruyama discretization for both $X_t$ and $Y_t$, we
obtain 

\begin{equation}
X_{t_{i+1}} = X_{t_i} + \mu(t_i,X_{t_i}) \Delta t_i + \sigma^T(t_i,X_{t_i})
\Delta W^i
\end{equation}
\begin{equation}
Y_{t_{i+1}} = Y_{t_i} - f\left( t_i,X_{t_i},Y_{t_i},\Pi_{t_i} \right) \Delta 
t_i + \Pi^T_{t_i} \sigma^T(t_i,X_{t_i}) \Delta W^i
\end{equation}
This can be used to time-step both $X_t$ and $Y_t$ forward. 

To time-step $Y_t$ backward, one needs to solve
\begin{equation} 
Y_{t_i} - f\left({t_i},X_{t_i},Y_{t_i},\Pi_{t_i} \right) \Delta {t_i} = 
Y_{t_{i+1}} - \Pi^T_{t_i}  \sigma^T(t_i,X_{t_i}) \Delta W^i 
\end{equation} 
for $Y_{t_i}$ (assuming $\Pi_{t_i}$ is given or otherwise determined, such as
by control or optimization). 

Set $F^B$ as the function that satisfies 
\begin{equation}
F^B(t,\Delta t,X,\Pi,R)-f(t,X,F^B(t,\Delta t,X,\Pi,R),\Pi) \Delta t = R
\end{equation}
then the exact backward step can be expressed as
\begin{equation}
 Y_{t_i} = F^B(t_i,\Delta t_i,X_{t_i},\Pi_{t_i}, Y_{t_{i+1}} - \Pi^T_{t_i} 
 \sigma^T(t_i,X_{t_i}) \Delta W^i)
\end{equation}

Instead of an exact solution, one can use Taylor expansion as in Liang, Xu, and
Li \cite{liang2019deep}.

The functional will be replaced by an appropriate time-discretized version 
such as 
\begin{equation}
J^F(\Pi_t,\Pi^{\mathsf{final}},\ldots) = E\left(\sum_0^{N-1}
{\mathsf{rc}}(t_i,{X}_{t_i},Y_{t_i},\Pi_{t_i}) \delta t_i +
{\mathsf{fc}}({X}_N,Y_N,\Pi^{\mathsf{final}})\right)
\end{equation}
and
\begin{equation}
J^B(\Pi_t,\Pi^{\mathsf{initial}},\ldots) = E\left(\sum_0^{N-1}
{\mathsf{rc}}(t_i,{X}_{t_i},Y_{t_i},\Pi_{t_i}) \Delta t_i +
{\mathsf{ic}}({X}_0,Y_0,\Pi^{\mathsf{initial}})\right)
\end{equation}
or similar. 

\subsection{Self-financing conditions}

Similarly to the derivation of the self-financing condition for the
time-continuous case, assuming a trading strategy vector $\Pi_{t_i}$ in force 
during the time step from $t_i$ to $t_{i+1}$, one obtains the part of the
time-discrete BSDE for the portfolio value $Y$ as follows
\begin{equation}
\mathsf{driftterm}(t_i,\Delta t_i, Y_{t_i},\Pi_{t_i}) + \Pi^T_{t_i}
\sigma_{LN}(t_i,X_{t_i}) \Delta W^i 
\end{equation}
where the drift term in the risk neutral case is 
\begin{equation}
\mathsf{driftterm}(t_i,\Delta t_i, Y_{t_i},\Pi_{t_i}) :=
(1+r(t) \Delta t_i) Y_{t_i}
\end{equation}
and in the differential rate case is (recall $\pi_0(t)=Y_t- \Pi \cdot 1$,
common growth rate under pricing measure for underliers is $r(t)$)
\begin{equation}
\mathsf{driftterm}(t_i,\Delta t_i, Y_{t_i},\Pi_{t_i}) :=
Y_{t_i} + \Delta t_i 
\left( r(t) \sum_{i=1}^N \pi_i(t_i) +  
\left\{
\begin{array}{lc} 
r^l(t) \pi_0(t_i) & \textrm{if } \pi_0(t_i) \geq 0 \\
r^b(t) \pi_0(t_i) & \textrm{else  } \\
\end{array}
\right.
\right)
\end{equation}

Dividends or fees for the underlying securities would contribute appropriate
terms.

So far we have not found time-discrete FBSDE with transaction costs handled by
deepBSDE methods. 

Transaction costs can be introduced in the following fashion: Value based
transaction costs incurred in the reallocation from $\Pi_{t_i}$ to
$\Pi_{t_{i+1}}$ contribute terms of the form (where $j$ indicates the
component/underlying security that incurs these transaction costs)
\begin{equation}
- \sum_{j=1}^n \lambda_j \left| \pi_j(t_{i+1}) - \pi_j(t_i) \right|^{q_j}
\end{equation}

Per share based transaction costs contribute terms of the form 
\begin{equation}
- \sum_{j=1}^n \lambda_j  \left|\frac{\pi_j(t_{i+1})}{X_j(t_{i+1})} -
\frac{\pi_j(t_i)}{X_j(t_{i})} \right|^{q_j}
\end{equation}

A fixed commission that is only charged when there is trading contributes terms
of the form (with $1_x$ the indicator function):
\begin{equation}
- \sum_{j=1}^n \lambda_j 1_{\pi_j(t_{i+1}) \neq \pi_j(t_i)}  
\end{equation}

Whatever the exact form is, the total $Y$ drift term can be written as a
function $f_{\Delta t}$ as follows in the time-discrete BSDE for $Y_i$:

\begin{equation}
Y_{t_{i+1}} = Y_{t_i} - f_{\Delta t} \left( t_i,\Delta t_i,
X_{t_i},X_{t_{i+1}},Y_{t_i},Y_{t_{i+1}},\Pi_{t_i},\Pi_{t_{i+1}} \right) 
+ \Pi^T_{t_i} \sigma^T(t_i,X_{t_i}) \Delta W^i
\end{equation}

Exact or approximate backward steps can be derived as in the previous
subsection. 

\subsection{Forward time-stepping methods}

In these methods, both the risk factors $X_{t_i}$ and the portfolio 
value $Y_{t_i}$ are time-stepped forwards. 
 
\subsubsection{Europeans}

Here, in the functional, there is typically no running cost, and the only term
is the $L_2$ distance or similar measure on the replication of the final values. 

After time-stepping the $Y_{t_i}$ forward, the functional given by the final
cost term is evaluated and gives the objective/loss function. 

Even if the functional would include running costs, one would carry along the
running cost together with the $X_{t_i}$ and $Y_{t_i}$ as, say $J_{t_i}$.

\subsubsection{Barriers}

We will discuss only the case of knock-out barriers with immediate rebate on
hitting the barrier. Other barrier option types can be treated similarly. 

The evolution of the $X_{t_i}$ is monitored and a barrier indicator
$\mathsf{Barrier}_{t_i}$ is introduced that turns from 0 to 1 when the evolution
of the $X_{t_i}$ breaches the barrier and otherwise remains constant (in
particular it stays at 1 even if $X$ should no longer breach the barrier at a
later time). We introduce additional state variables $t^B_{t_j}$, $X^B_{t_j}$,
and $Y^B_{t_j}$ that record the time $t_i$, the $X_{t_i}$, and the $Y_{t_i}$ at
barrier breach or maturity (whatever comes first) at a time $t_j$ at or after
the breach/maturity.
Depending on the value of $t^B_{t_N}$ (maturity or not), the final cost is the
$L_2$ norm of the difference (or other risk measure) between $Y^B_{t_N}$ and the
appropriate payoff $g(X_{t_N})$ or $g_b(t^B_{t_N},X^B_{t_N})$.

This is a new method. We are currently implementing this method with promising
results and expect to publish details of the method and results soon. 

\subsubsection{Exercise Opportunities}

First, one determines the exercise strategy or the holding value, for instance
by starting forward or backward method with random initial risk factor values at
the potential exercise time.

Given $\mathsf{hv}(X_{t_E})$, at time $t_E$, we will check
$g_E(X_{t_E})>\mathsf{hv}(X_{t_E}) $.\footnote{If the exercise strategy function
is given directly, we check whether it indicates exercise at $t_E$ and proceed
in the same fashion.} If that is true, $t_E$ will be marked as exercise time
$t^E$, $X_{t_E}$ as the risk factor values $X^E$ at exercise, and $Y_{t_E}$ as
the value $Y^E$ at exercise, otherwise, $t^E$ will be maturity and $X^E$ and
$Y^E$ are the corresponding values at maturity. The final cost will be
determined as

\begin{equation}
\mathsf{fc}(X^E,Y^E) = || Y^E - G(t^E,X^E) ||^2 
\end{equation}
where 
\begin{equation}
G(t^E,X^E) = 
\left\{
\begin{array}{lc} 
g(X^E) & \textrm{if } T^E=T \\
g_E(X^E) & \textrm{if  } T^E=t_E \\
\end{array}
\right.
\end{equation}

In the context of FBSDE, this is a new method. We are currently implementing
this method with promising results and expect to publish details of the method and results soon. 

\subsection{Backward time-stepping methods}

In these methods, the risk factors $X_{t_i}$ are time-stepped forwards 
and the portfolio value $Y_{t_i}$ is time-stepped backwards.

\subsubsection{Europeans}
Here, in the functional, there is typically no running cost, and the only term
is the variance of the initial values of $Y$, $Y_0$ (for fixed $X_0$) or the
$L_2$ distance of the $Y_0$ from a to-be-determined function
$Y^{\mathsf{init}}(X_0)$, which can be evaluated once $Y$ has been time-stepped
back to the initial time. 

Even if the functional would include running costs, one would carry along the
running cost together with the $X_{t_i}$ and $Y_{t_i}$ as, say $J_{t_i}$, and
then add the initial term as described above to compute the total cost/total
value of the functional. 

\subsubsection{Barriers}
We will disccuss only the case of knock-out barriers with immediate rebate on
hitting the barrier. Other barrier option types can be treated similarly. 

The
evolution of the $X_{t_i}$ is monitored and a barrier indicator
$\mathsf{InBarrier}_{t_i}$ is introduced that will be equal to 1 only at the
times when the $X_{t_i}$ are in the barrier region and the barrier is active and
be equal to 0 at all other times.

After $Y$ has been time-stepped backward from $t_{i+1}$ to $t_i$ as in the
European case, $Y$ will be overwritten with the value of the knocked-in rebate
$g_B(t_i,X_{t_i})$ if $\mathsf{InBarrier}_{t_i}$ is 1 and will be unchanged
otherwise. In this way, the correct value of $Y$ in the barrier is always
enforced. 

The initial cost does not change. If there is running cost, the running cost
term is reset to zero (or the running cost corresponding to the knocked-in
instrument).

This is a new method. We are currently implementing this method with promising
results and expect to publish details of the method and results soon. 

\subsubsection{Exercise Opportunities}

After $Y$ has been time-stepped backwards until $t_E$, the holding value based
exercise strategy $g_E(X_{t_E})>\mathsf{hv}(X_{t_E}) $ or the directly given
exercise strategy $\mathsf{es}(X_{t_E})$ will be checked. If the used strategy
indicates exercise, $Y$ will be overwritten by the exercise value $g_E(X_{t_E})$. 

This is a new method. We are currently implementing this method with promising
results and expect to publish details of the method and results soon. 

Alternatively, without determining an exercise strategy or holding value, $Y$
could be overwritten with the exercise value $g_E(X_{t_E})$ if the exercise
value is larger than the backward time-stepped portfolio value. This exercise
strategy is not adapted and cannot be applied in general unless the future is
known (and will lead to noisy results with clairvoyance/foresight bias). 
This approach has been used by Wang et al \cite{wang2018deep} for Bermudan
swaptions in the LMM model. 

One can also determine some exercise strategy based on a mini-batch or other
approximation of the dynamics, within the optimization. 
This has been proposed and applied in Liang, Xu, and
Li \cite{liang2019deep} for Bermudan and callable products. 

\section{Solving time-discrete FBSDE stochastic control problems by Deep Neural
Networks and Deep Learning}

Given the time-discrete appropriately time-stepped FBSDE and the appropriate
stochastic control functional (written here to cover both forward and backward
methods):
\begin{equation}
J(\Pi_t,\Pi^{\mathsf{initial}},\Pi^{\mathsf{final}}) =
E\left(\sum_0^{N-1}
{\mathsf{rc}}(t_i,{X}_{t_i},Y_{t_i},\Pi_{t_i}) \Delta t_i +
{\mathsf{fc}}({X}_N,Y_N,\Pi^{\mathsf{final}}) + 
{\mathsf{ic}}({X}_0,Y_0,\Pi^{\mathsf{initial}})
\right)
\end{equation}
we would need to know the laws of all the $X_{t_i}$ and $Y_{t_i}$ and we would
need to integrate against all the appropriate joint probability density
functions to compute the exact value of this functional. 
Instead, whenever one wants to estimate the value of the functional given a
particular control, one does so by a Monte-Carlo type estimate, sampling the
discrete-time processes $X_{t_i}$ and $Y_{t_i}$ (and other additional state
variable processes as appropriate) along a certain number of realizations. 

In general, we will iteratively improve the parameters of the control by
gradient descent methods such as stochastic gradient descent or mini-batch
methods; or more involved stochastic optimization methods such as Adam
algorithm. 

If the functional contains exercise strategies to be determined from a number of
sample realizations or a variance or other things that need to be determine
from a number of paths or from one or several batches, one has to specify how to
determine the appropriate batch-level quantities or expressions. 

We now assume that the strategy process $\Pi_{t_i}$ is given by deep neural
networks of appropriate architectures taking as input $X_{t_i}$ but potentially 
also other inputs (and if we model all $\Pi_t$ at the same time rather than
separately for different $t_i$, $t_i$ would be one such extra input). 
These strategy processes can also take some state or output prepared from the
strategy at an earlier time, such as in the case of Long short-term memory
(LSTM) or similar architectures. In the case of random $X_0$, we also assume
$Y^{\mathsf{init}}(X_0)$ to be given by a deep neural network. 

\subsection{Deep Neural Networks}

In the literature, there are many different architectures given. In the context
of deep BSDE methods, Chan-Wai-Nam, Mikael, and Warin
\cite{chan2019machine} presents several choices. 

One of the most straightforward settings is a fully connected 
feedforward deep network. Assume we want to approximate a function from
$R^{d_1}$ to $R^{d_2}$ and assume we have intermediate layer sizes 
$m_0=d_1$, $m_1$, $m_2$,\ldots, $m_L=d_2$. One way to
describe such is to define them as a composition of affine transformations 
$A_l(x)$ and component-wise applications of activation functions 
\begin{equation}
N(x;\Theta) = \rho_L \circ A_L \circ \rho_{L-1} \circ A_{L-1} \circ
\ldots \circ \rho_1 \circ A_1 \circ \rho_0 (x) 
\end{equation}
where $\rho_i(x)=(\rho_i(x_1),..,\rho_i(x_{m_i}))$ with activation functions
$\rho_i$ such as the sigmoid, the ReLu, the Elu, tanh, swish, mish, etc. and
$A_l$ being maps from $R^{m_{l-1}}$ to $R^{m_l}$ with appropriate matrices $A_l(x) = W_l x
+ b_l$. The first and/or the last activation function can be the identity. All
the parameters contained in the $W_l$ and $b_l$ (and, if appropriate, any
parameters for the $\rho_l$) will be collected into the parameter collection
$\Theta$.

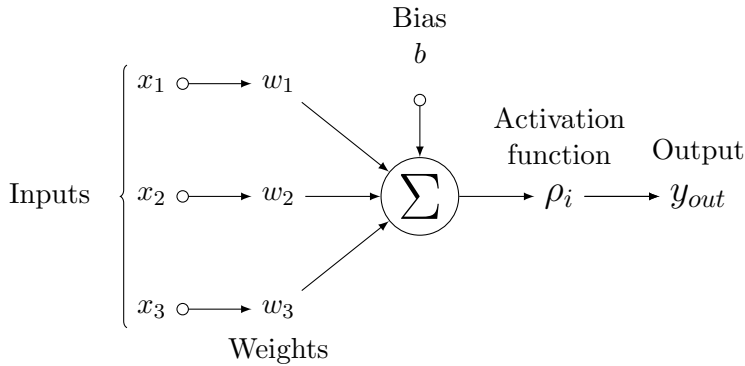
\begin{figure}

\begin{center}
    \begin{tikzpicture}[
        init/.style={ 
             draw, 
             circle, 
             inner sep=2pt,
             font=\Huge,
             join = by -latex
        },
        squa/.style={ 
            font=\Large,
            join = by -latex
        }
    ]
        
        \begin{scope}[start chain=1]
            \node[on chain=1] at (0,1.5cm)  (x1) {$x_1$};
            \node[on chain=1,join=by o-latex] (w1) {$w_1$};
        \end{scope}
        
        \begin{scope}[start chain=2]
            \node[on chain=2] (x2) {$x_2$};
            \node[on chain=2,join=by o-latex] {$w_2$};
            \node[on chain=2,init] (sigma) {$\displaystyle\Sigma$};
            \node[on chain=2,squa,label=above:{\parbox{2cm}{\centering
            Activation\\ function}}]   {$\rho_i$}; \node[on
            chain=2,squa,label=above:Output,join=by -latex] {$y_{out}$};
        \end{scope}
        
        \begin{scope}[start chain=3]
            \node[on chain=3] at (0,-1.5cm) 
            (x3) {$x_3$};
            \node[on chain=3,label=below:Weights,join=by o-latex]
            (w3) {$w_3$};
        \end{scope}
        
        \node[label=above:\parbox{2cm}{\centering Bias \\ $b$}] at (sigma|-w1) (b) {};
        
        \draw[-latex] (w1) -- (sigma);
        \draw[-latex] (w3) -- (sigma);
        \draw[o-latex] (b) -- (sigma);
        
        \draw[decorate,decoration={brace,mirror}] (x1.north west) -- node[left=10pt] {Inputs} (x3.south west);
        
    \end{tikzpicture}
    
\end{center}
\caption{Example neuron in a (D)NN \label{dnn_neuron}}
\end{figure}

In diagrams, this might look like the following:
A single neuron with three inputs (and therefore three weights) and one bias and
an activation function $\rho_i$ is shown in figure \ref{dnn_neuron}.

A standard feedforward dense deep neural network as used in our tests for an
input dimension of 4 for $\pi_i(X_i)$ in which no parameters are shared is
shown in figure \ref{pidnn} (each of the neurons implements an operation as
shown in the previous diagram).

\def\layersep{2.5cm}
\def\twolayersep{5.0cm}
\def\threelayersep{7.5cm}

\begin{figure}
\begin{center}
\begin{tikzpicture}[shorten >=1pt,->,draw=black!50, node distance=\layersep]
    \tikzstyle{every pin edge}=[<-,shorten <=1pt]
    \tikzstyle{neuron}=[circle,fill=black!25,minimum size=17pt,inner sep=0pt]
    \tikzstyle{input neuron}=[neuron, fill=green!50];
    \tikzstyle{output neuron}=[neuron, fill=red!50];
    \tikzstyle{hidden neuron}=[neuron, fill=blue!50];
    \tikzstyle{annot} = [text width=4em, text centered];

    \foreach \name / \y in {1,...,4}
			\node[input neuron, pin=left:$x^i_{\name}$] (I-\name) at (0,-\y) {};

    \foreach \name / \y in {1,...,14}
        \path[yshift=5cm]
            node[hidden neuron] (H1-\name) at (\layersep,-\y cm) {};

    \foreach \name / \y in {1,...,14}
        \path[yshift=5cm]
            node[hidden neuron] (H2-\name) at (\twolayersep,-\y cm) {};

    \foreach \name / \y in {1,...,4}
        \path
            node[output neuron, pin=right:$\pi^i_{\name}$] (O-\name) at
            (\threelayersep,-\y cm) {};

    \foreach \source in {1,...,4}
        \foreach \dest in {1,...,14}
            \path (I-\source) edge (H1-\dest);

    \foreach \source in {1,...,14}
        \foreach \dest in {1,...,14}
            \path (H1-\source) edge (H2-\dest);

    \foreach \source in {1,...,14}
    	\foreach \dest in {1,...,4}
        	\path (H2-\source) edge (O-\dest);

    \node[annot,above of=H1-1, node distance=1cm] (hl1) {Hidden layer 1};
    \node[annot,above of=H2-1, node distance=1cm] (hl2) {Hidden layer 2};
    \node[annot,left of=hl1] {Input layer};
    \node[annot,right of=hl2] {Output layer};
\end{tikzpicture}
\end{center}
\caption{Example DNN for the portfolio function $\pi^i(X_i)$  for an
four-dimensional case.\label{pidnn}}
\end{figure}
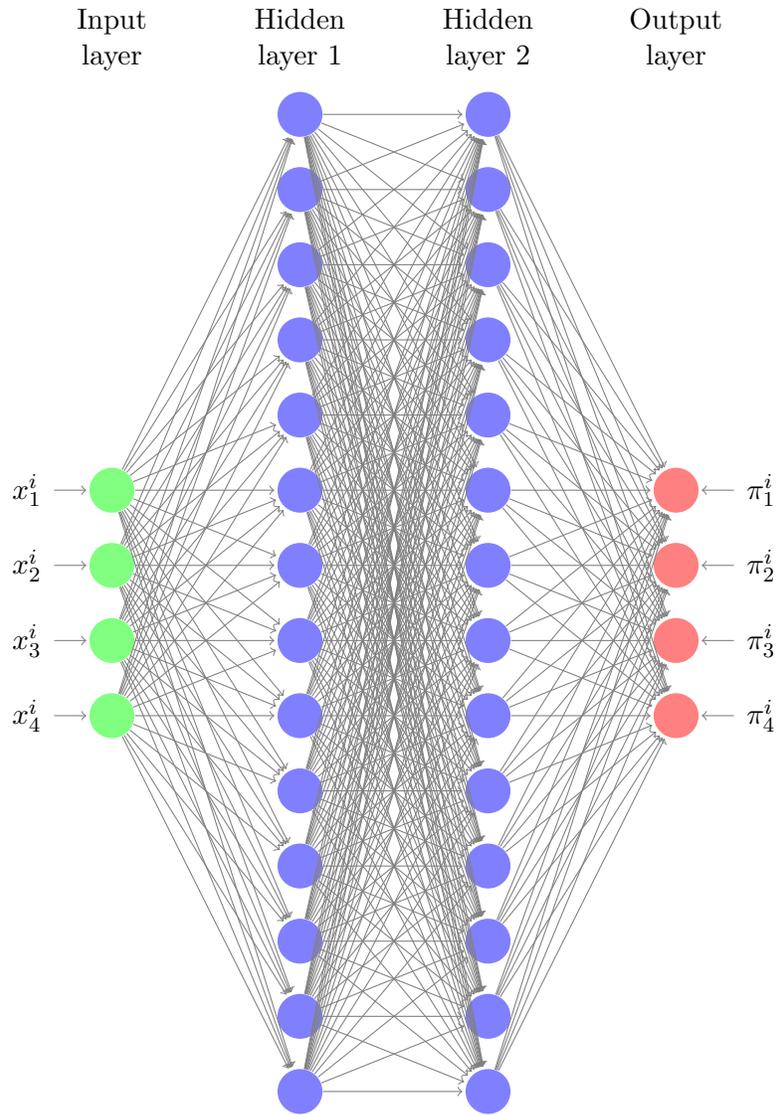

A similar network but with a single output would be used for
$Y^{\mathsf{init}}(X_0)$.

In general, stochastic optimization methods/deep neural network modeling
perform best if the inputs to any network (layer) are in an appropriate
non-dimensional scale, ranging from say -1 to 1 (centered around zero with a
relatively small width in the single digits). If the range of the input can be
controlled or determined ahead of time, one can just subtract off the center and
divide by the (half-)width (''prescaling''). If the range is not known or
previous layers or changes in the input stream are driving inputs to undesirable
ranges, one can apply batch-normalization. That means that one uses the
information from a mini-batch to normalize the input to a layer (or output) and
learn the appropriate parameters. The quantities computed from mini-batches in
training are replaced by moving averages or population averages during
inference. Batch-normalization was introduced by Ioffe and Szegedy
\cite{ioffe2015batch} and is described in the following:

The batch normalization transform/layer, as used in the training of the
networks, has two parameters $\gamma$ and $\beta$ that are learned during the
training. ($\epsilon$ is a non-trainable parameter that prevents division
by zero or very small numbers.) On a mini-batch of length $B$, it consist of the
following operations:

\begin{eqnarray}
\mu_B & = & \frac {1}{B} \sum_{b=1}^B x_b \\
\sigma^2_B & = & \frac {1} {B} \sum_{b=1}^B (x_b - \mu_B)^2 \\
\hat{x}_b &= & \frac{x_b - \mu_B} {\sqrt{\sigma^2_B + \epsilon}} \\
y_b & = & \gamma \hat{x}_b + \beta \\ 
\end{eqnarray}

The original input or output $x_b$ to or from the layer is replaced by the so
computed $y_b$. This layer/transform can be added before and/or after each
layer, as desired.
It might be enough to put it at the beginning of the network in some cases. 

When evaluating the results of the network on new mini-batches for inference and
similar, $\mu_B$ and $\sigma^2_B$ are replaced by moving averages or averages
or values otherwise computed, say $\mu_I$ and $\sigma^2_I$. For instance, the
original paper suggests to compute them over multiple (training) mini-batches.
However they are computed, the batch-normalization layer from training is
replaced by
\begin{equation}
y = \frac{\gamma}{\sqrt{\sigma^2_I + \epsilon}} \cdot x + \left(\beta -
\frac {\gamma \mu_I}{\sqrt{\sigma^2_I + \epsilon}}\right)
\end{equation}
during evaluation/inference. 

\subsection{Deep Learning Stochastic Optimization Approaches}

We will denote by $\Theta$ the collection of all the parameters in all of the
controls ($\Pi_t$,$\Pi^{\mathsf{initial}}$,$\Pi^{\mathsf{final}}$, and
$Y^{\mathsf{init}}(X_0)$, as far as they are used in the method and setup under
consideration.) We will denote the reparametrization of the functional 
$J(\Pi_t,\Pi^{\mathsf{initial}},\Pi^{\mathsf{final}},\ldots)$ with the
parameter collection $\Theta$ as $J^{\Pi}(\Theta;\ldots)$.

Pure stochastic gradient descent (SGD) would be 
\begin{equation}
 \Theta \gets  \Theta - \eta  \nabla_\Theta
  J^{\Pi}(\Theta;X(\omega),Y(\omega),\ldots) 
\end{equation}
and a (mini-)batch approach would be
\begin{equation}
 \Theta \gets  \Theta  - \eta  \frac{1}{B} \sum_{b=1}^{B} \nabla_\Theta
  J^{\Pi}(\Theta;X(\omega_b),Y(\omega_b),\ldots) 
\end{equation}
where $X(\omega)$ and $Y(\omega)$ are realizations of $X_t$ and $Y_t$ (and also
any other state variables are assumed to be simulated to obtain realizations of
them, as necessary). 
To explain some of the later approaches more concisely, we
will denote any of these approximate gradients by \[
G_n(\Theta_n,X(\omega),Y(\omega),\ldots) \]

Pure gradient descent methods sometimes change directions too often and gradient
descent directions in certain geometries will lead to oscillatory behavior. One
possible mitigation is to not change directions completely but to mix previous
direction and the current gradient together, as in momentum methods (so called
since momentum keeps you going). It can be applied to any approximation of the
gradient.

The new descent direction $\Delta \Theta$ is determined by 
\begin{eqnarray}
\Delta \Theta_{n+1} & = & \alpha \Delta \Theta_{n+1} - \eta G_n(\Theta_n,\ldots)
\\
\Theta_{n+1} & =& \Theta_n + \Delta \Theta_{n} 
\end{eqnarray}

Another commonly used algorithm is called Adam (Adaptive Moment Estimation)
suggested by Kingma and Ba \cite{kingma2014adam}:

The Adam optimizer requires a learning rate $\alpha$, two parameters
$\beta_1$ and $\beta_2$ in $[0,1)$ that determine the exponential decay rates
for the moment estimates, and one parameter $\epsilon$ that prevents division
by zero.
The two (biased) moments $m_0$ and $v_0$ are initialized to 0.

The Adam algorithm then updates the parameters according to 

\begin{eqnarray}
g_{n} & = & G_n(\Theta_n,X(\omega),Y(\omega)) \\
m_n & = & \beta_1 m_{n-1} + (1-\beta_1) g_n \\
v_n & = & \beta_2 v_{n-1} + (1-\beta_2) g^2_n \\
\hat{m}_n &= & \frac{m_n} {1-\beta^n_1} \\
\hat{v}_n & = & \frac{v_n} {1-\beta^n_2} \\
\Theta_{n+1} & =& \Theta_{n} - \alpha \frac
{\hat{m}_n}{\sqrt{\hat{v}_n}+\epsilon}
 \end{eqnarray}

New optimization methods and variants and/or combinations of older optimization
methods are steadily coming out and are being published at a steady rate, such
as rectified Adam, Lookahead, and Ranger, which we will not discuss here. However,
any such promising algorithms are typically implemented in TensorFlow, Keras, or
PyTorch and can therefore be relatively easily applied.

\section{Examples}

We will present some simple example from our current work. We will pick an
one-dimensional Black-Scholes setting since for this setting we have analytical
solutions and the results can be easily visualized. (Multi-dimensional
extensions have been obtained for geometric basket options and other situations
but visualization would be more of a challenge in such cases.)

We consider the one-dimensional Black-Scholes model with constant drift and
short rate of 0.06 and constant volatility of 20\% (0.2). We either start the
underlier at a spot of 120 or vary it uniformly between 70 and 170. We consider
a combination of a long call at 120 and two short calls at 150 (which leads to
delta of varying sign so that the differential rates model would lead to
non-linear pricing). Both calls have maturity of half a year, 0.5. We discretize
time with 50 time steps. 

We use mini-batches of size 512. Our networks have 4 layers of sizes
1,11,11, and 1. We prescale with given center
and width in underlier. 
We used ELU as an activation function for all layers except the output
layer, for which we used identity.

Learning rate is 1e-3 and we use Adam with Tensor Flow standard parameters. We
will compute loss functions for validation for randomly chosen mini-batches of
the same size. The loss function for validation is therefore computed as an MC
sample (and we can try to get an idea for the distribution by repeatedly
computing it given different mini-batches/samples). We run 20000 mini-batches.

We train separate networks $\pi_i$ mostly, but present one example where the
$\pi_i$ networks share parameters (and have an additional $t$ input).

\subsection{Forward methods}

\begin{figure}
\begin{center}
\includegraphics[width=75mm]{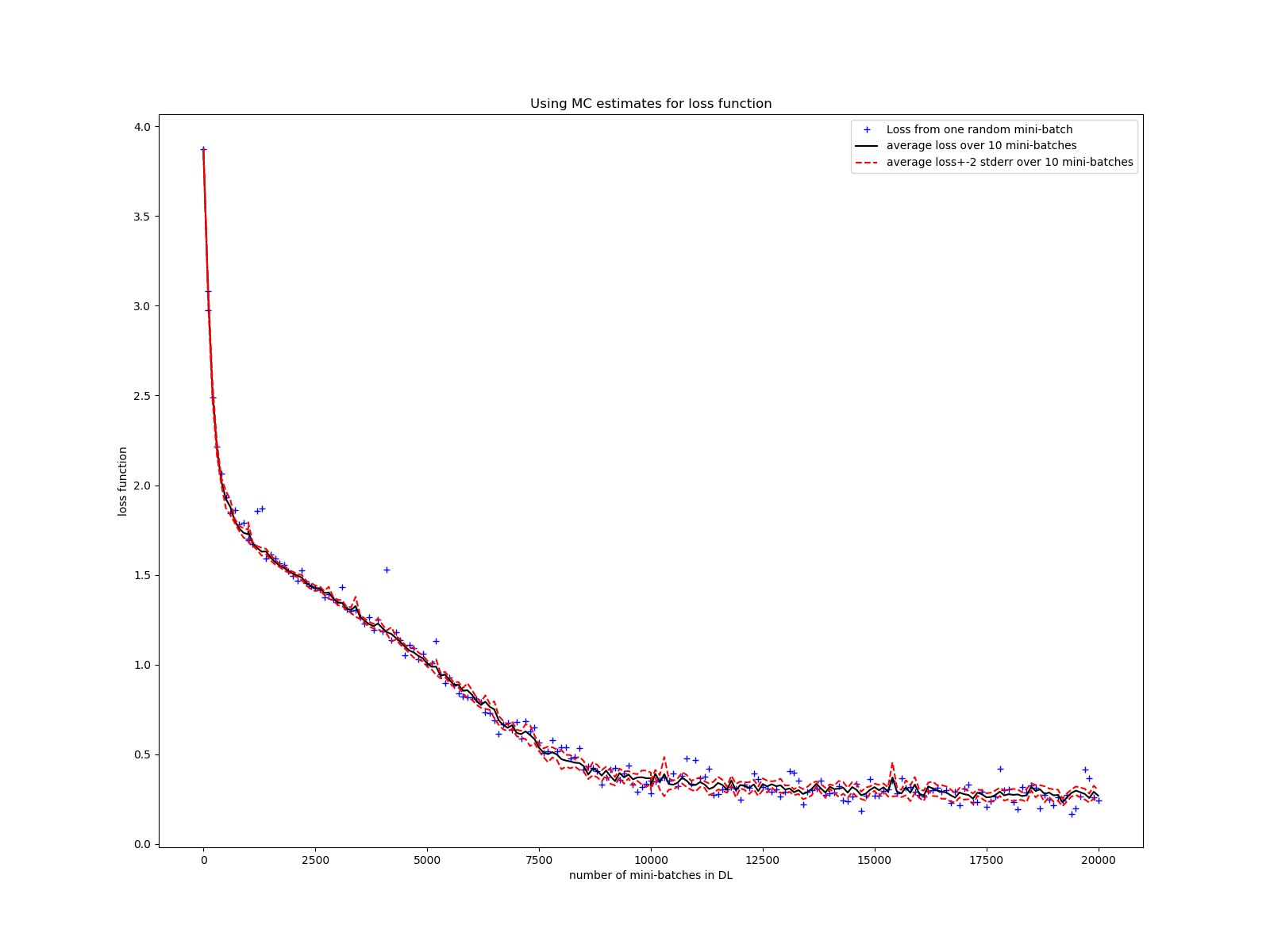}
\includegraphics[width=75mm]{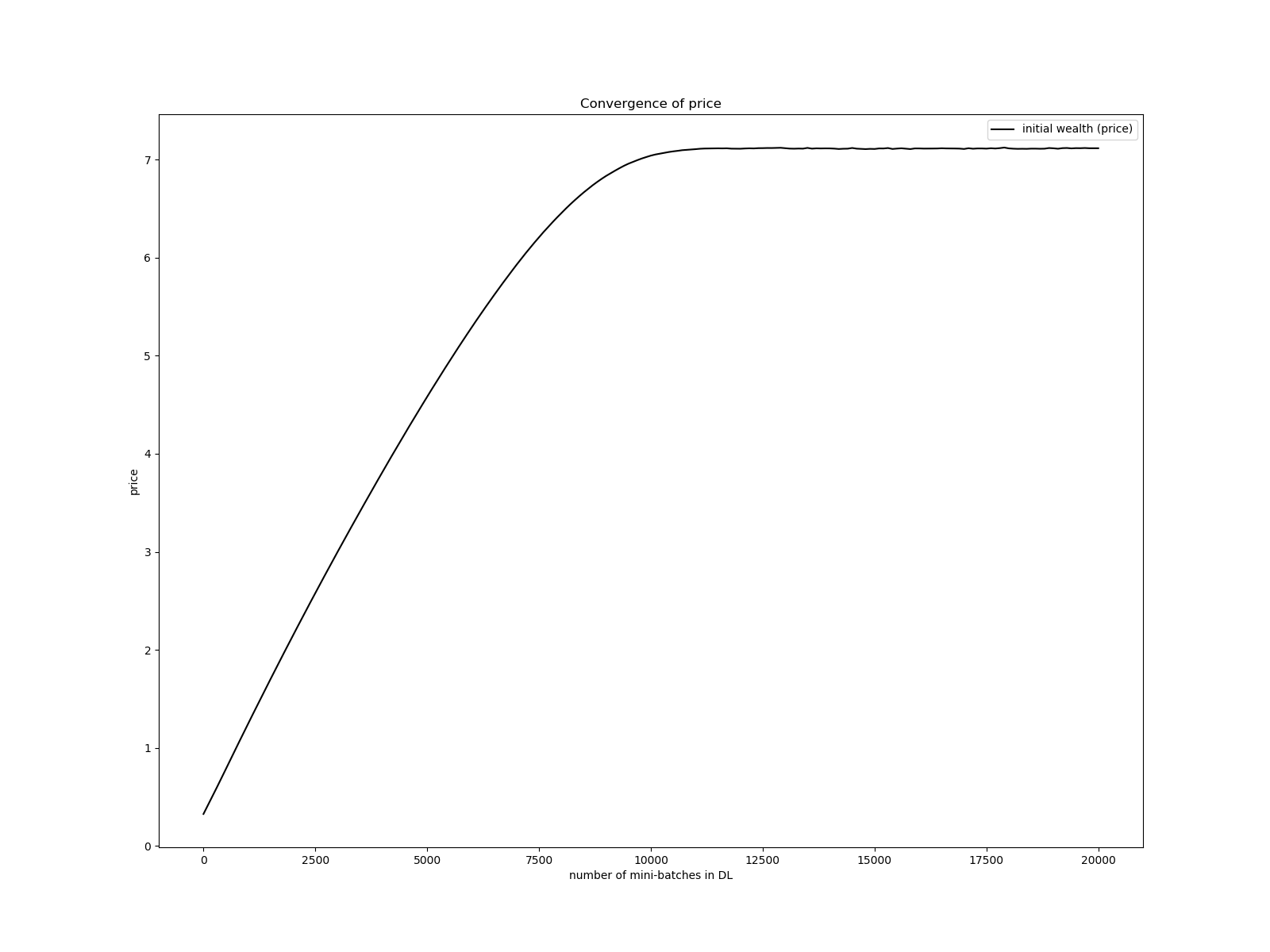}%
\includegraphics[width=75mm]{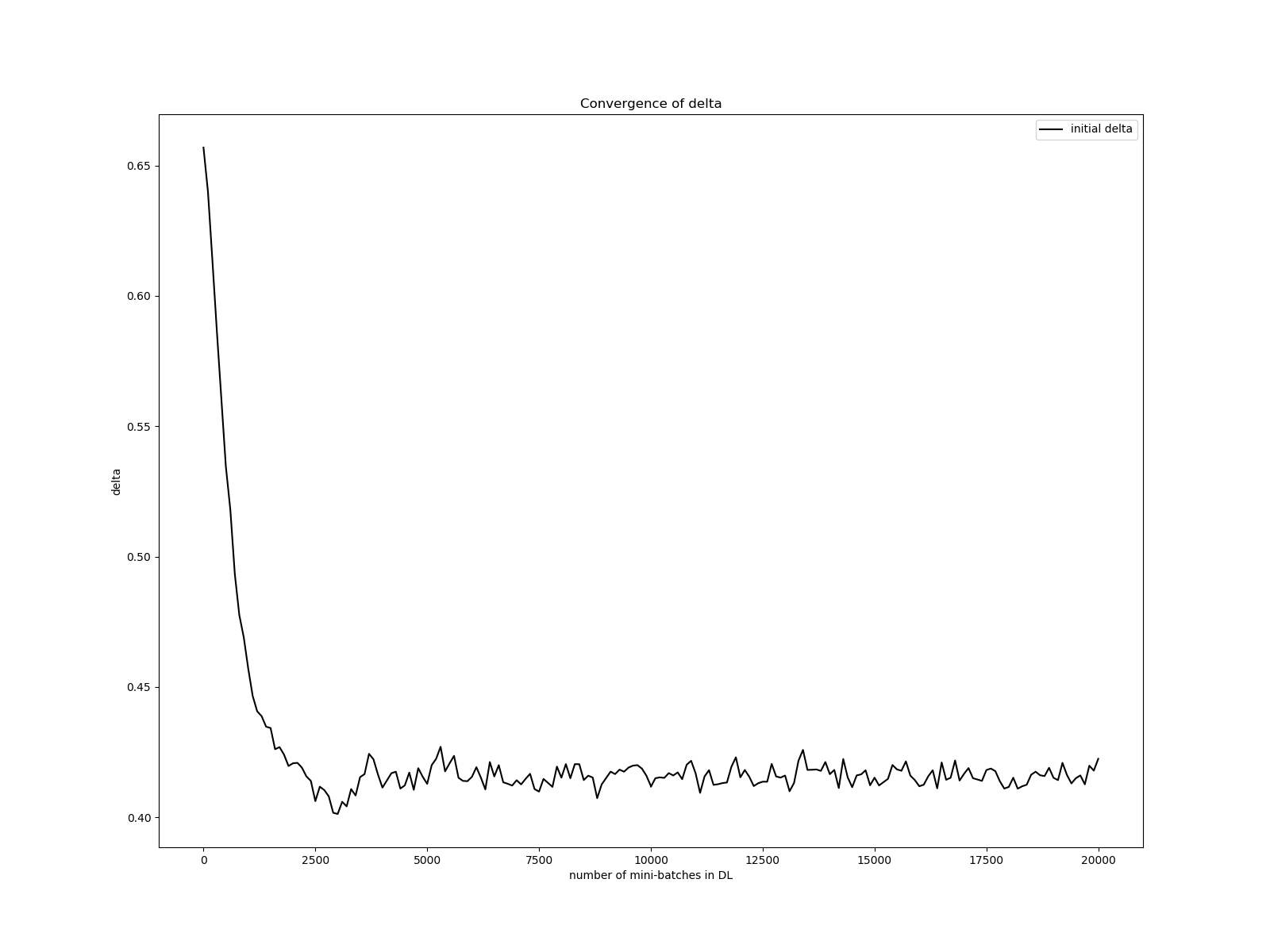}
\end{center}
\caption{Forward method with fixed initial risk factor values. 
Upper: loss
function/functional over mini-batch number.
Lower-left: convergence of price. Lower-right: convergence of
delta.\label{FwdFixedIVResults}}
\end{figure}

First, some results for the variant with fixed initial risk factors. Figure
\ref{FwdFixedIVResults} shows that the loss functional decays quite quickly as a
function of number of mini-batches run and how price and delta at fixed initial risk
factor (``spot'') converge up to good accuracy. 

\begin{figure}
\begin{center}
\includegraphics[width=75mm]{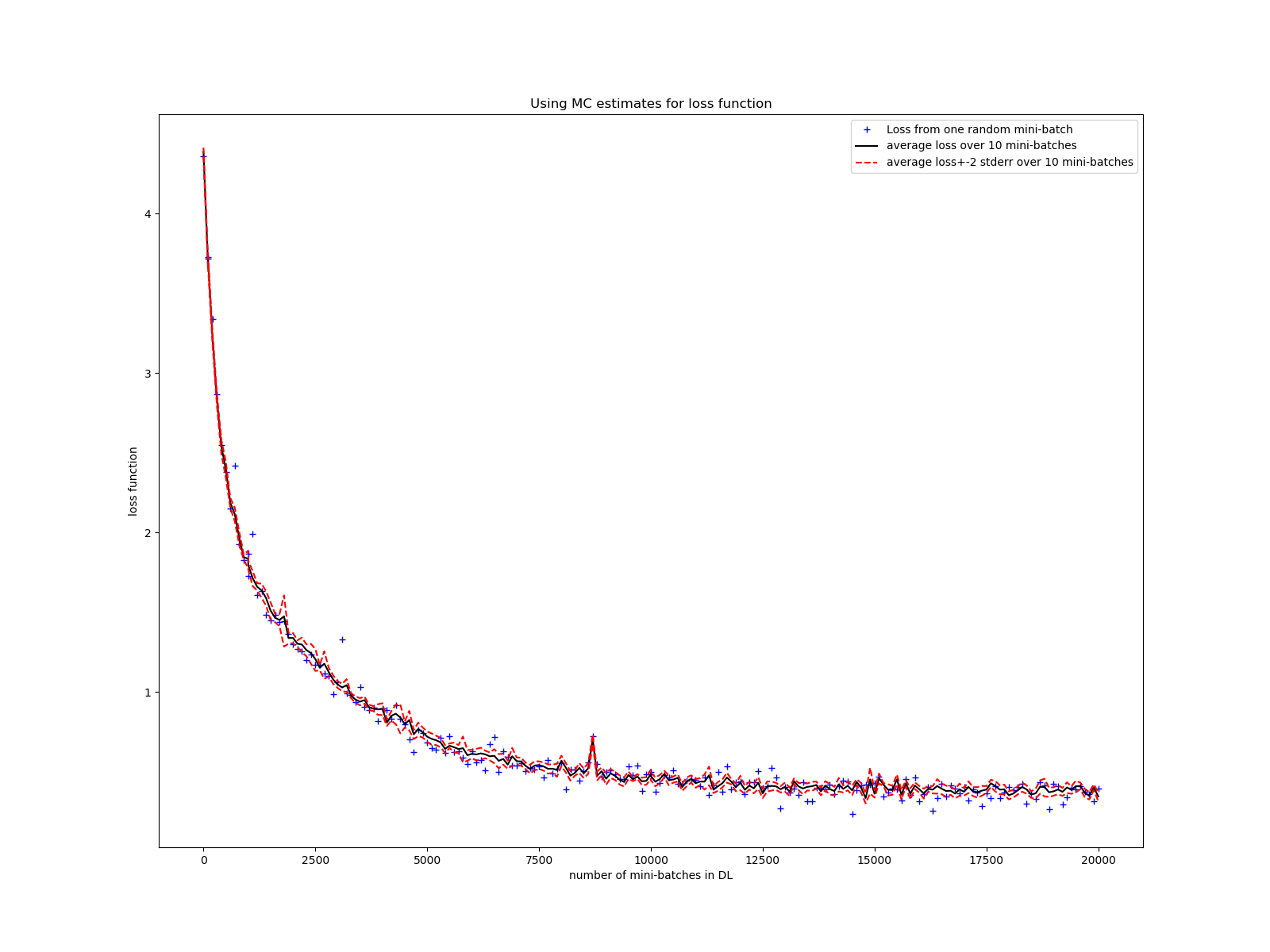}%
\includegraphics[width=75mm]{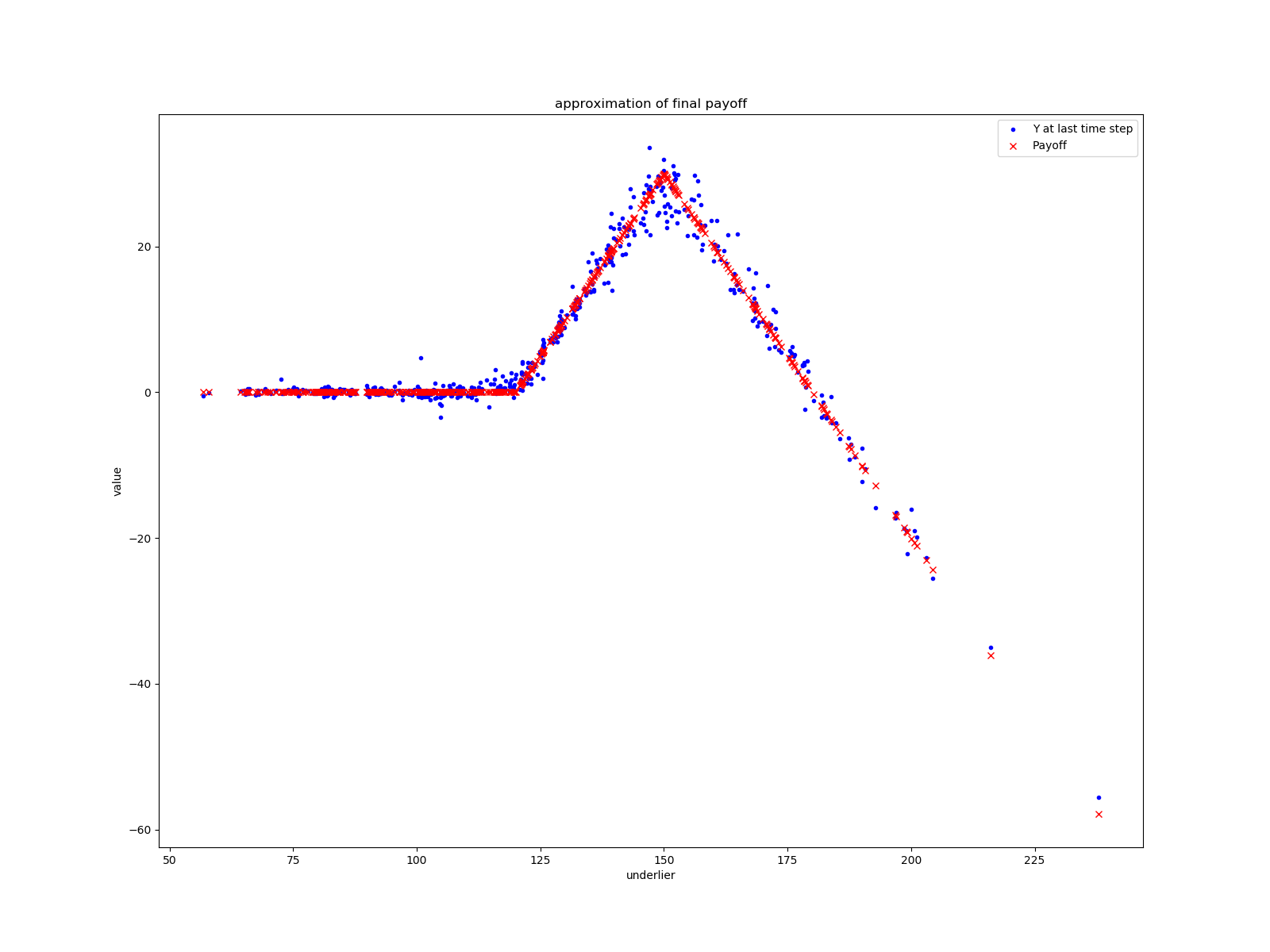}
\includegraphics[width=75mm]{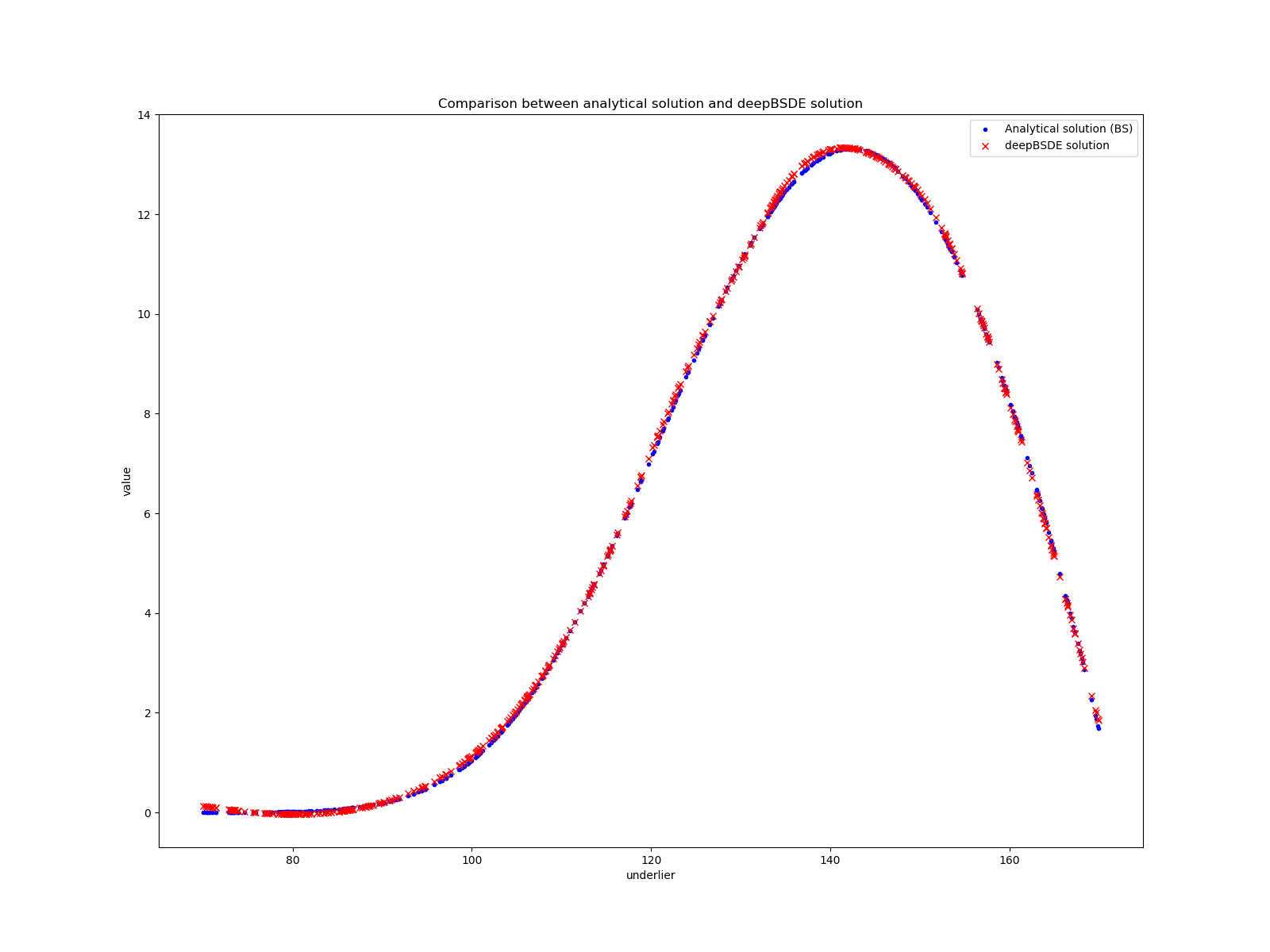}%
\includegraphics[width=75mm]{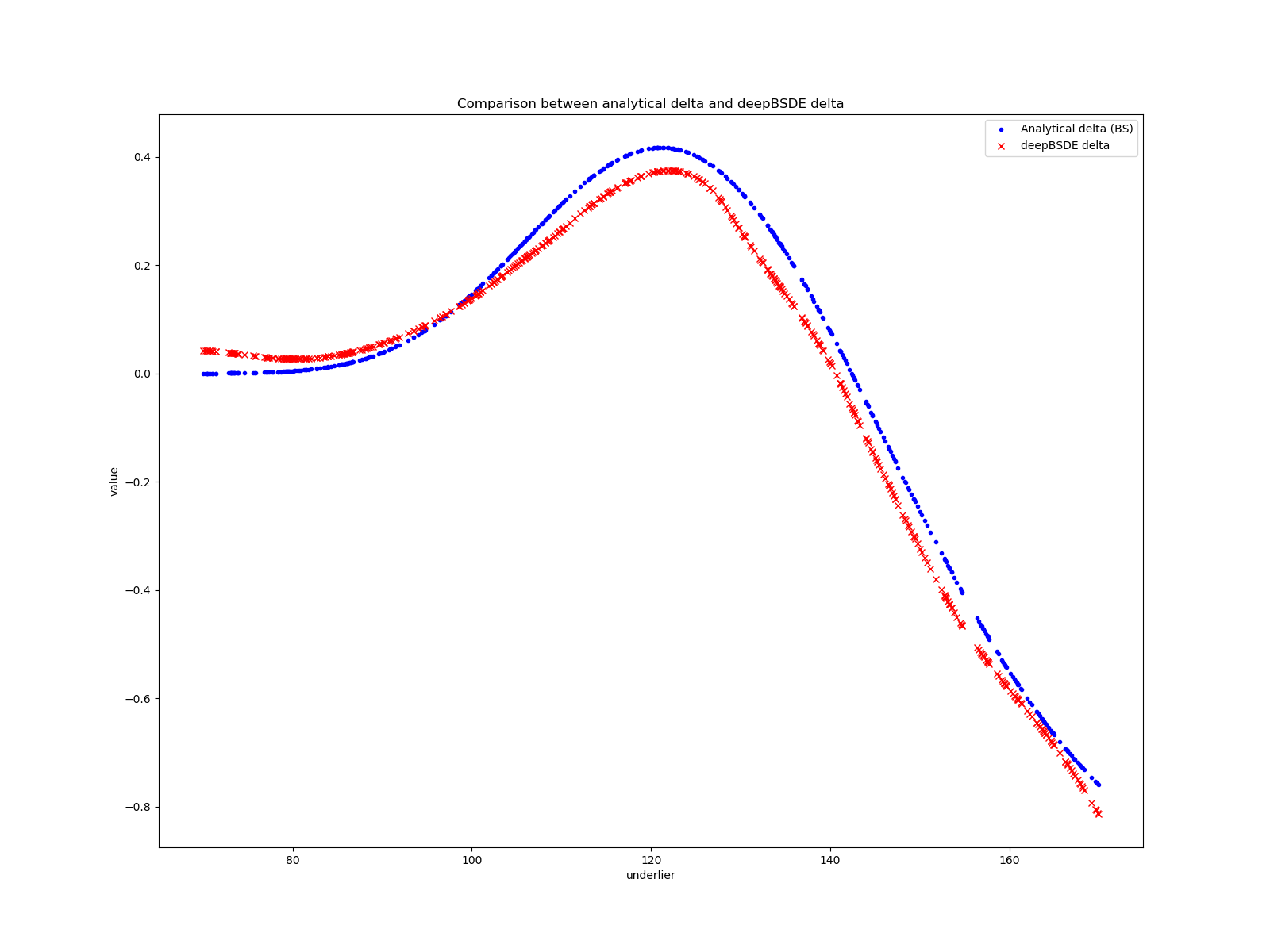}
\includegraphics[width=75mm]{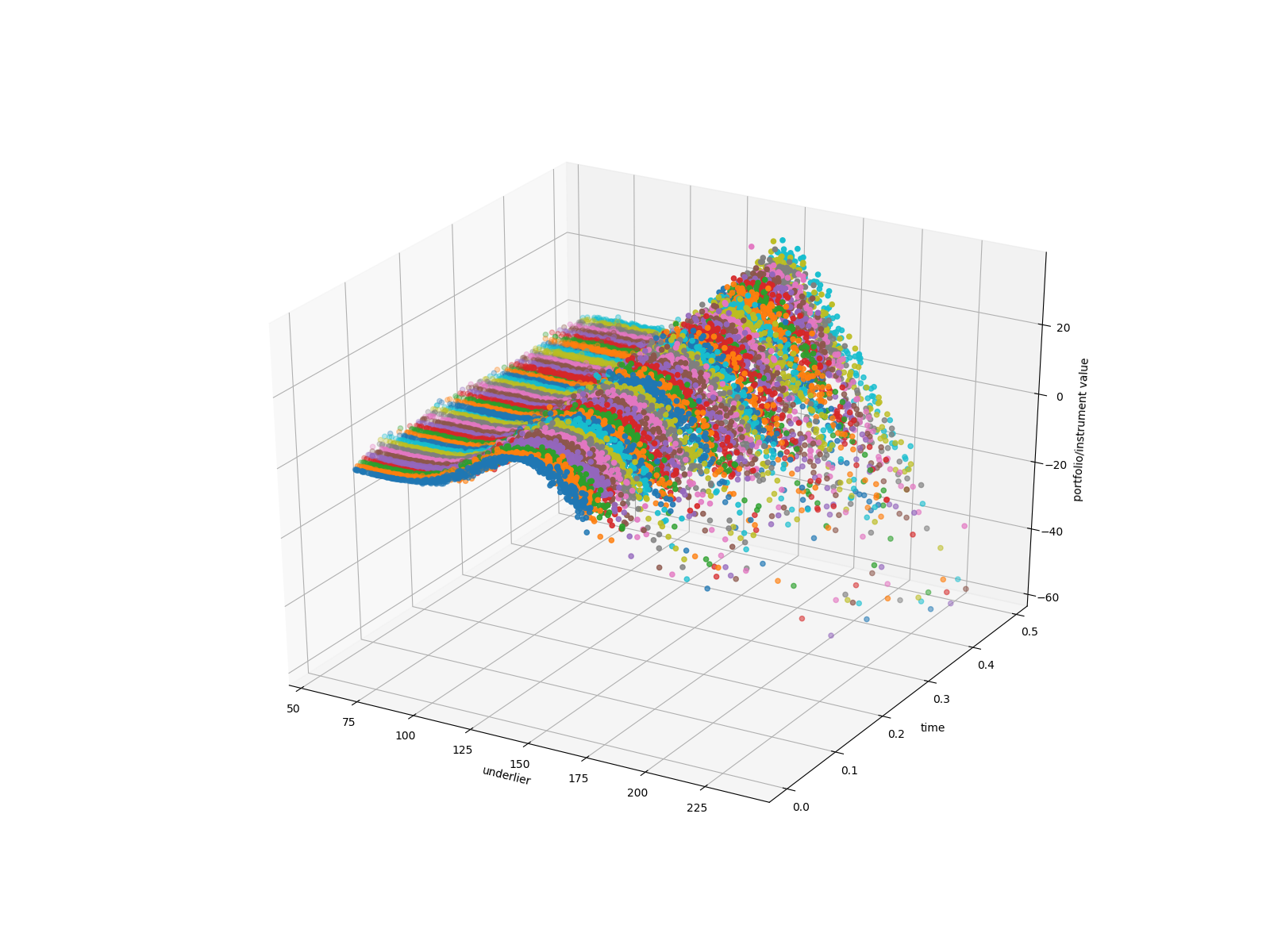}%
\includegraphics[width=75mm]{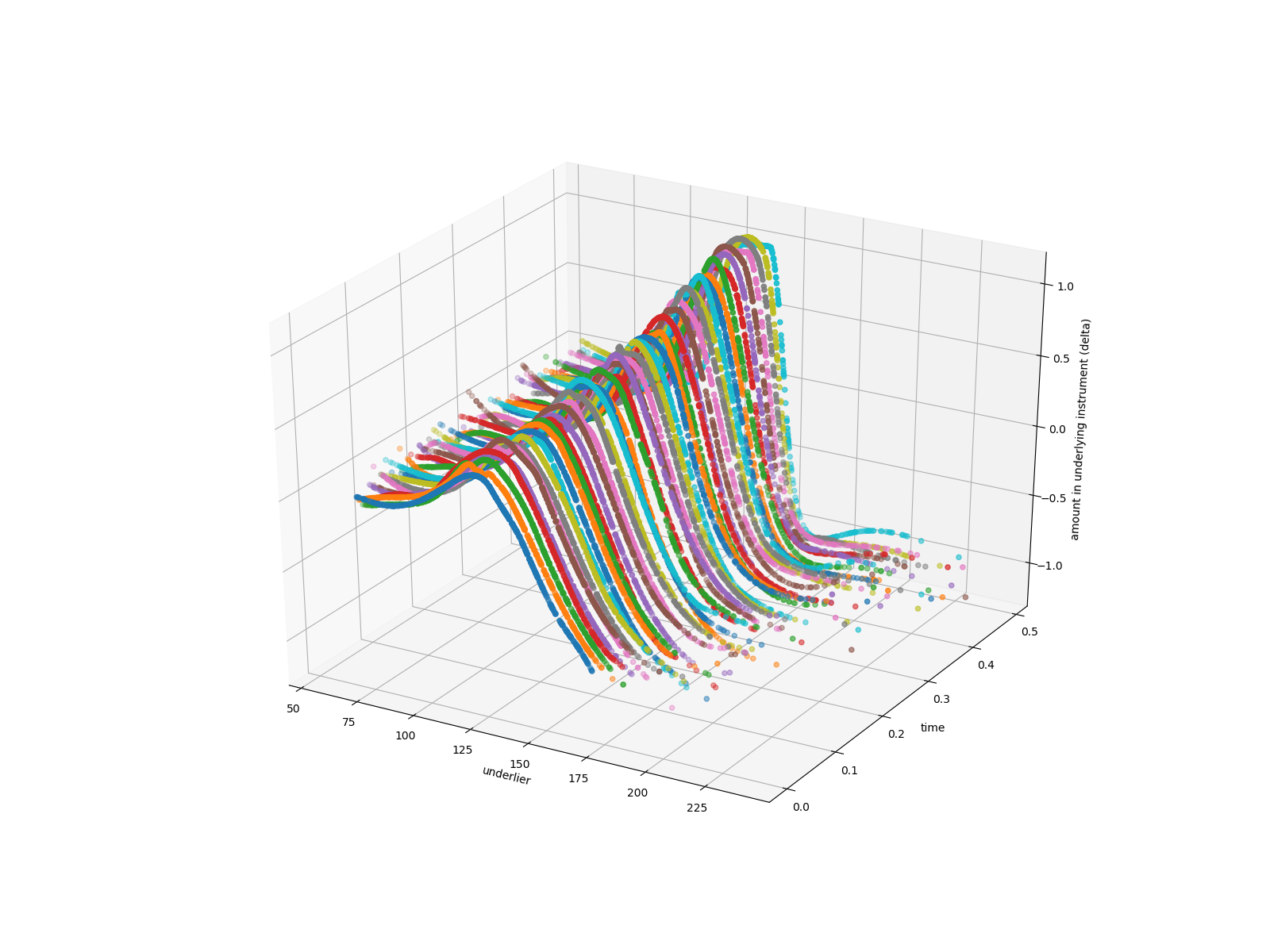}
\end{center}
\caption{Forward method with random initial risk factor values. Upper-left:
Loss function/functional over mini-batch number.
Upper-right: Final Payoff match at 20000. Middle-left: comparison of initial
$Y$ network and analytical solution. Middle-right: comparison of initial
portfolio delta and analytical delta. Lower-left: scatter plot of $Y$.
Lower-right: scatter plot of $\Pi$.\label{FwdRandomIVResults} }
\end{figure}

Now for the variant with random initial risk factors: Figure
\ref{FwdRandomIVResults} shows that the loss functional decays quite quickly for
this case also, that the final payoff is well replicated, that the analytical
solution is well approximated, the delta is quite well approximated as well
(considering that we only hedge at discrete times and not continuously), and
that the $Y$ surface and the portfolio functions surface looks well-defined and
smooth. 

\subsection{Backward methods}

\begin{figure}
\begin{center}
\includegraphics[width=75mm]{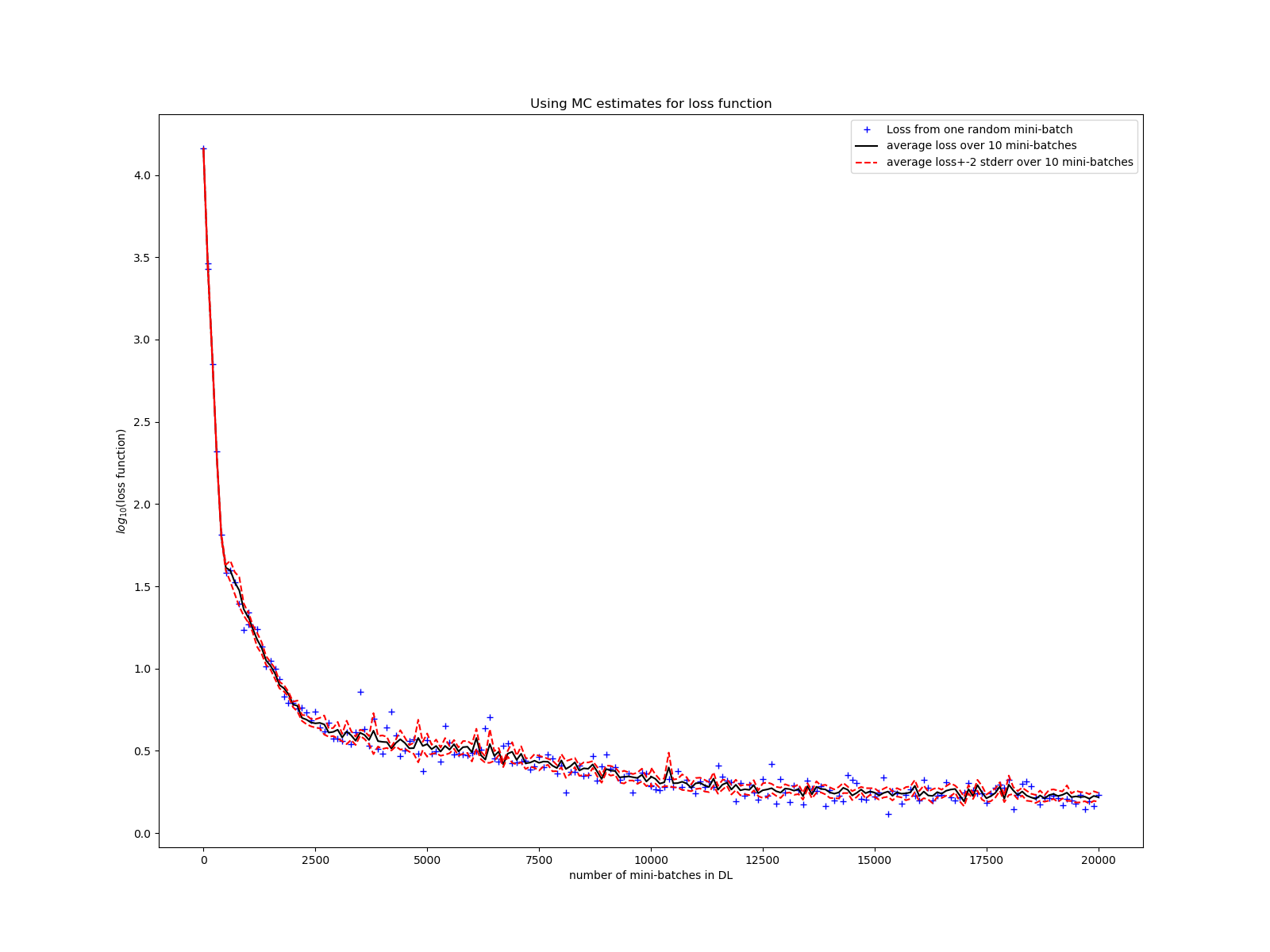}
\includegraphics[width=75mm]{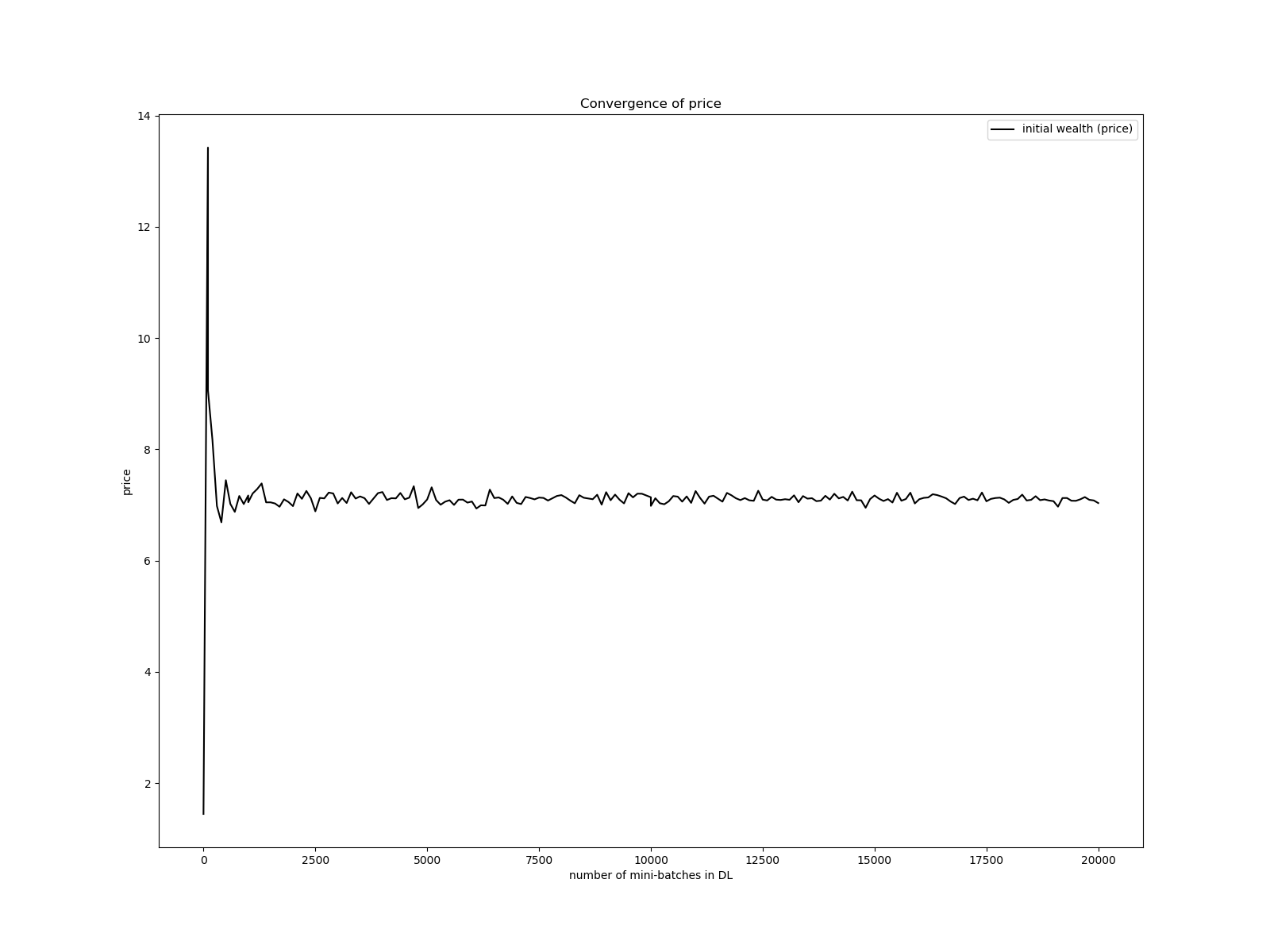}%
\includegraphics[width=75mm]{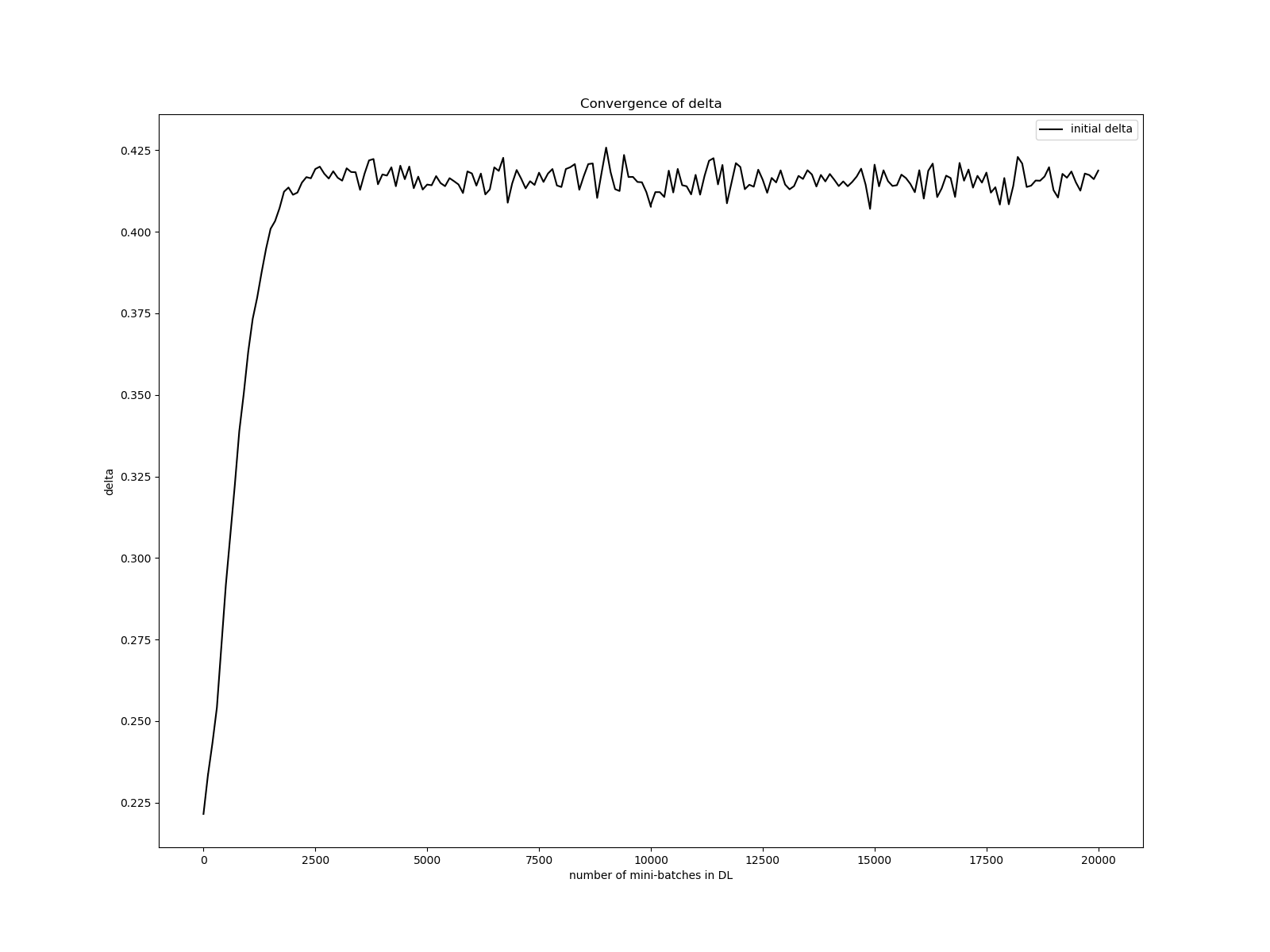}
\end{center}
\caption{Backward method with fixed initial risk factor values. 
Upper: loss
function/functional over mini-batch number.
Lower-left: convergence of price. Lower-right: convergence of
delta.\label{BkwdFixedIVResults}}
\end{figure}

First, some results for the variant with fixed initial risk factors. Figure
\ref{BkwdFixedIVResults} shows that the loss functional decays quite quickly as
a function of number of mini-batches run and how price and delta at fixed initial risk
factor (``spot'') converge up to good accuracy. (In this case, we actually
trained $\Pi$ networks with shared parameters - which means that the network
also has $t$ as an input.) We see that the backward method seems to converge
faster than the forward method in this case. 

\begin{figure}
\begin{center}
\includegraphics[width=75mm]{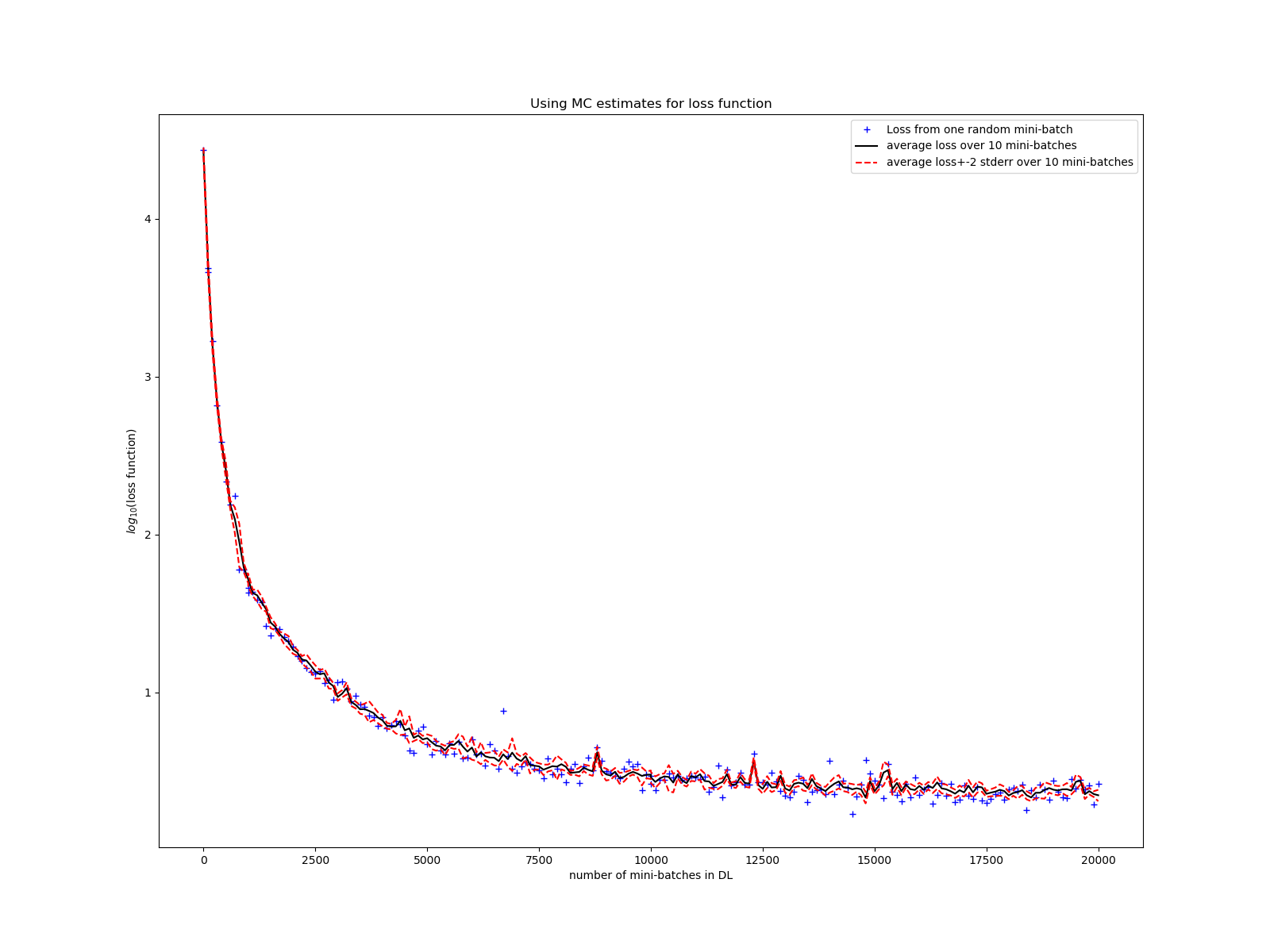}
\includegraphics[width=75mm]{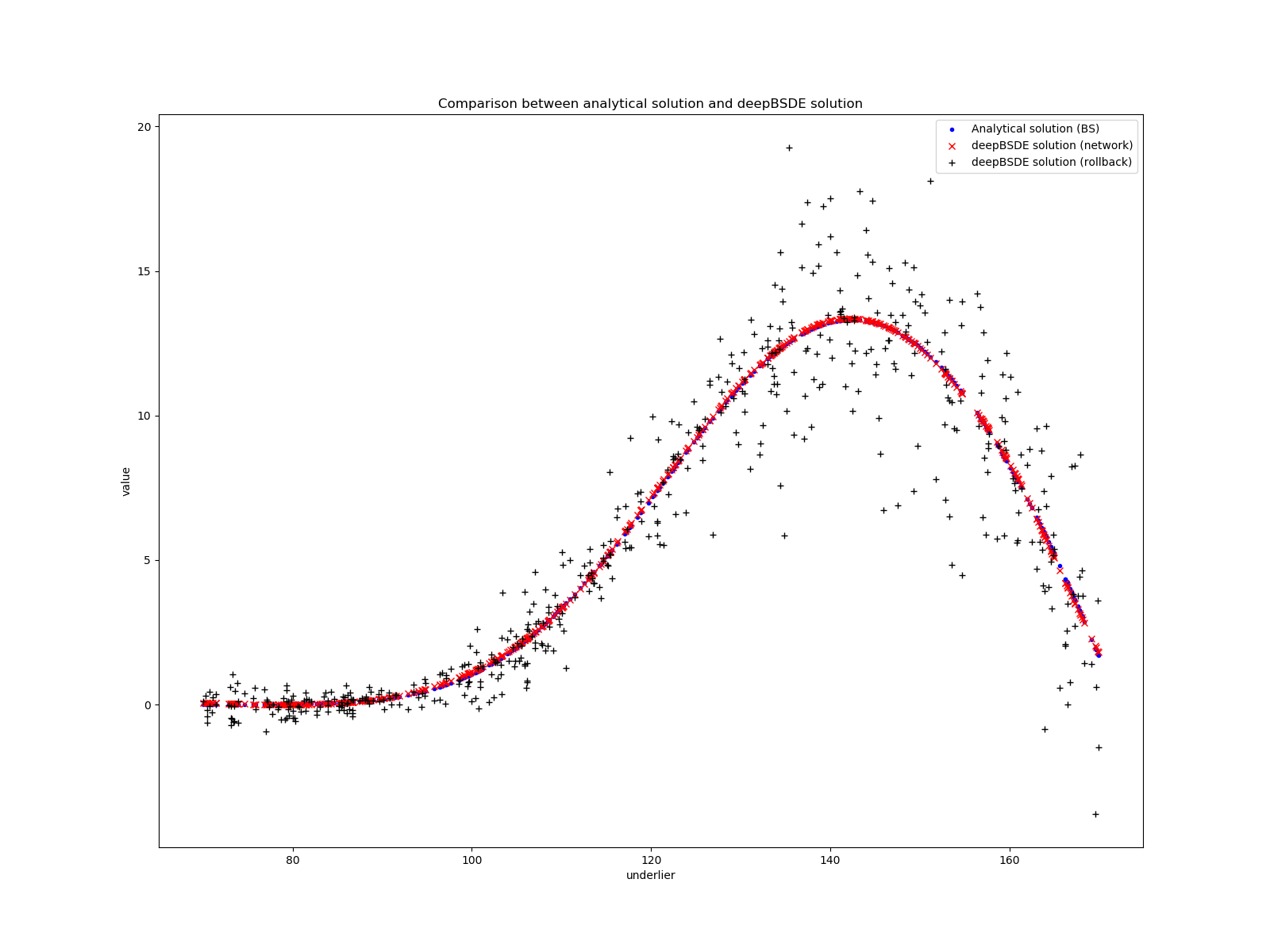}%
\includegraphics[width=75mm]{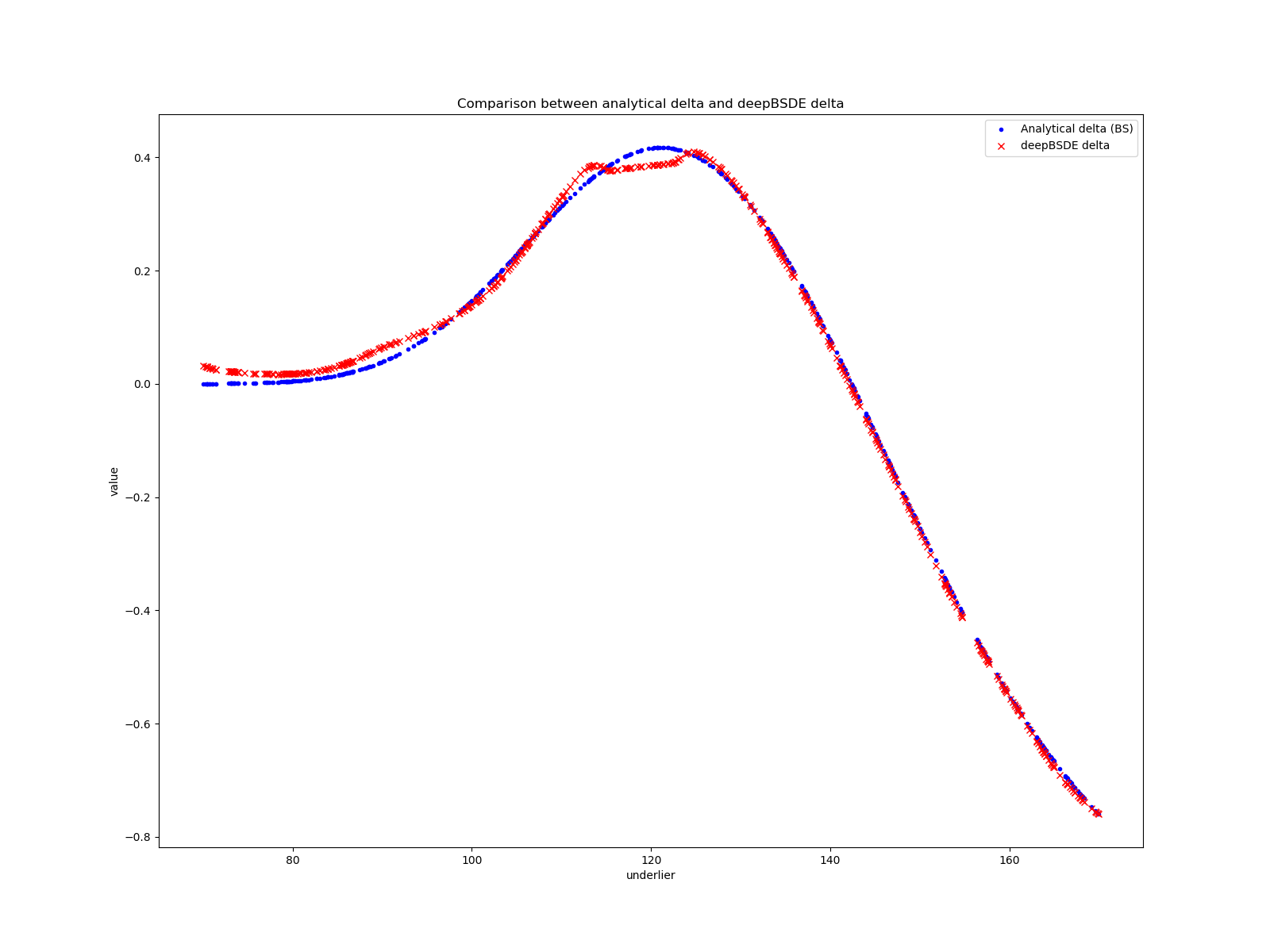}
\includegraphics[width=75mm]{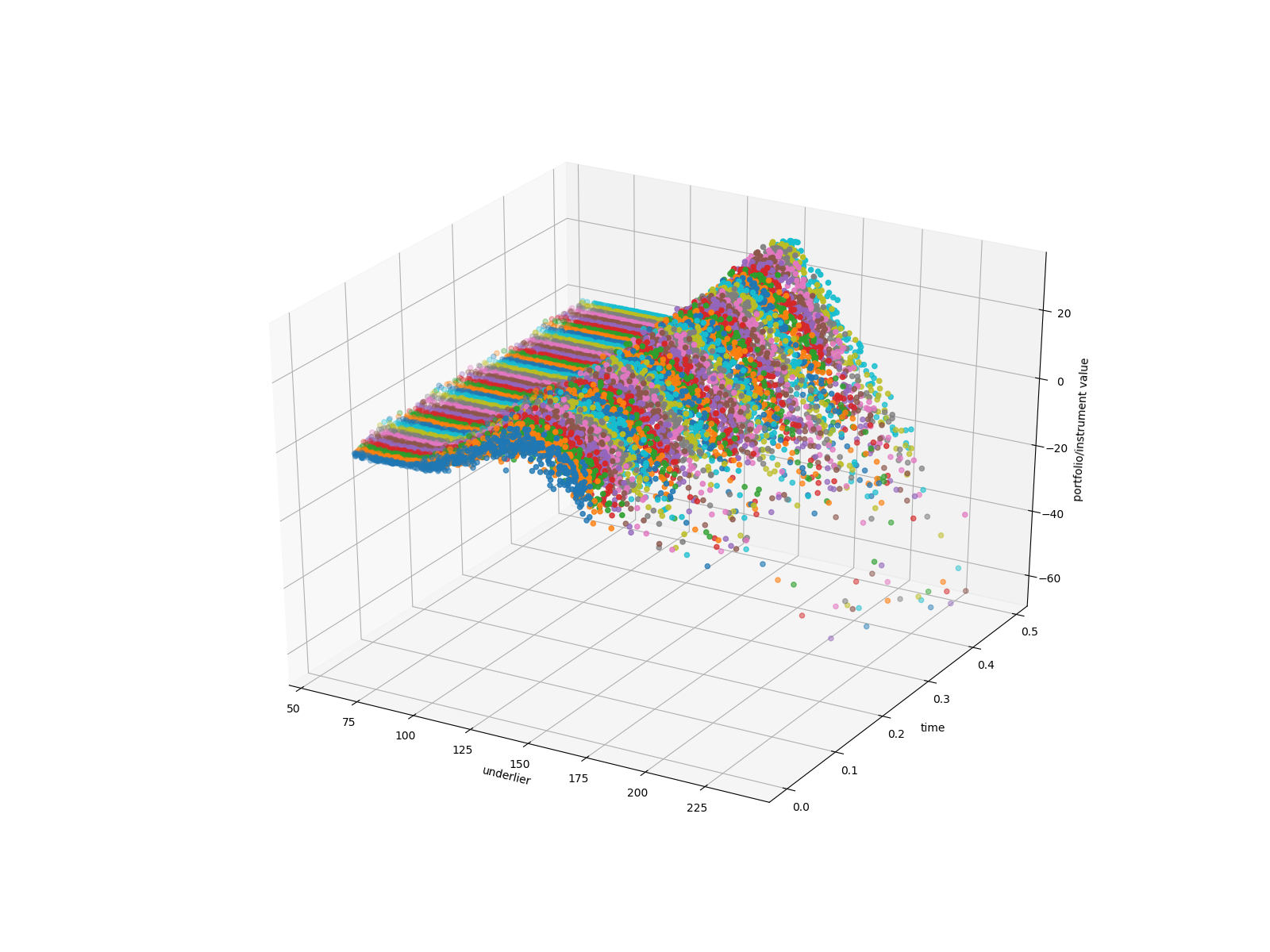}%
\includegraphics[width=75mm]{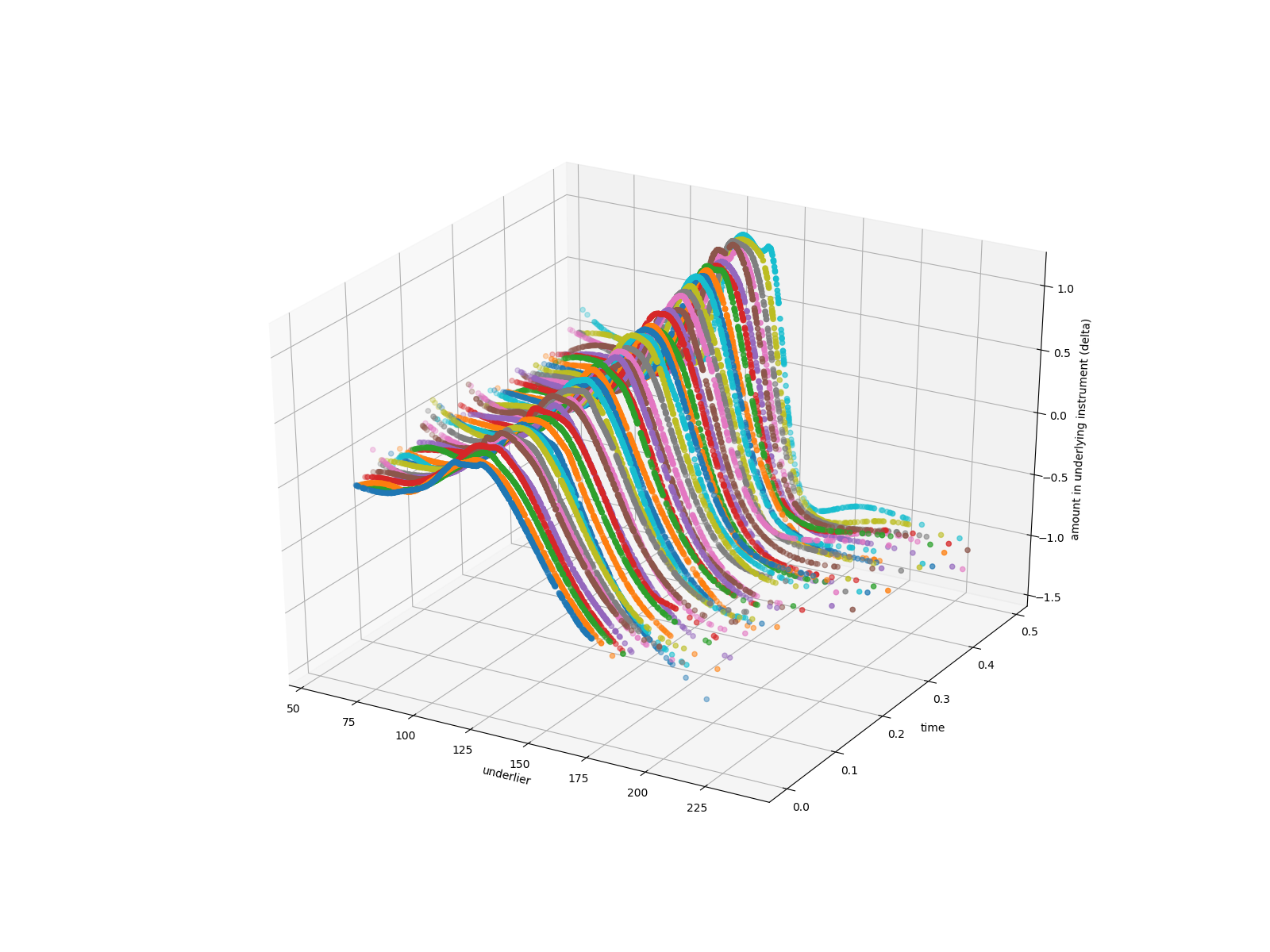}
\end{center}
\caption{Backward method with random initial risk factor values. 
Upper:
Loss function/functional over mini-batch number.
Middle-left: comparison of initial
$Y$ network, rolled-back $Y_0$, and analytical solution. 
Middle-right:
comparison of initial portfolio delta and analytical delta. 
Lower-left: scatter plot of $Y$.
Lower-right: scatter plot of $\Pi$.\label{BkwdRandomIVResults} }
\end{figure}

Now for the variant with random initial risk factors: Figure
\ref{BkwdRandomIVResults} shows that the loss functional decays quite quickly
for this case also, that the
analytical initial solution is well approximated and that the rolled-back $Y_0$
are concentrated around the initial solution and network, the delta is quite
well approximated as well 
(considering that we only hedge at discrete times and not continuously), and that the $Y$ surface and the portfolio functions surface looks well-defined and
smooth. 

\section{Conclusion}

We demonstrated how a wide variety of modeling approaches in quantitative
finance for European, Barrier, and Bermudan option pricing can be solved
through deep learning optimization approaches for the forward-backward
stochastic differential equation (FBSDE) formulations where the forward and the
backward SDE are time-stepped to simulate pathwise values and showed examples
for European option pricing for both the forward and backward approaches. 

\bibliographystyle{alpha}
\bibliography{../FirstOverviewPaperDraft/reviewRefs}

\newcommand{\etalchar}[1]{$^{#1}$}
\begin{thebibliography}{CWNMW19}

\bibitem[CWNMW19]{chan2019machine}
Quentin Chan-Wai-Nam, Joseph Mikael, and Xavier Warin.
\newblock Machine learning for semi linear {P}{D}{E}s.
\newblock {\em Journal of Scientific Computing}, 79(3):1667--1712, 2019.
\newblock arXiv:1809.07609.

\bibitem[EHJ17]{weinan2017deep}
Weinan E, Jiequn Han, and Arnulf Jentzen.
\newblock Deep learning-based numerical methods for high-dimensional parabolic
  partial differential equations and backward stochastic differential
  equations.
\newblock {\em Communications in Mathematics and Statistics}, 5(4):349--380,
  2017.
\newblock arXiv:1706.04702.

\bibitem[EKPQ97]{el1997backward}
Nicole El~Karoui, Shige Peng, and Marie~Claire Quenez.
\newblock Backward stochastic differential equations in finance.
\newblock {\em Mathematical finance}, 7(1):1--71, 1997.
\newblock Also available on semanticscholar.org.

\bibitem[GMKS19]{gonon2019asset}
Lukas Gonon, Johannes Muhle-Karbe, and Xiaofei Shi.
\newblock Asset pricing with general transaction costs: Theory and numerics.
\newblock {\em arXiv preprint arXiv:1905.05027}, 2019.

\bibitem[HJE18]{han2018solving}
Jiequn Han, Arnulf Jentzen, and Weinan E.
\newblock Solving high-dimensional partial differential equations using deep
  learning.
\newblock {\em Proceedings of the National Academy of Sciences},
  115(34):8505--8510, 2018.

\bibitem[IS15]{ioffe2015batch}
Sergey Ioffe and Christian Szegedy.
\newblock Batch normalization: {A}ccelerating deep network training by reducing
  internal covariate shift.
\newblock {\em arXiv preprint arXiv:1502.03167}, 2015.

\bibitem[KB14]{kingma2014adam}
Diederik~P Kingma and Jimmy Ba.
\newblock Adam: A method for stochastic optimization.
\newblock {\em arXiv preprint arXiv:1412.6980}, 2014.

\bibitem[LXL19]{liang2019deep}
Jian Liang, Zhe Xu, and Peter Li.
\newblock Deep learning-based least square forward-backward stochastic
  differential equation solver for high-dimensional derivative pricing.
\newblock {\em arXiv preprint arXiv:1907.10578}, 2019.
\newblock Also available at SSRN: https://ssrn.com/abstract=3381794 or
  http://dx.doi.org/10.2139/ssrn.3381794.

\bibitem[Per10]{perkowski2010}
Nicolas Perkowski.
\newblock Backward {S}tochastic {D}ifferential {E}equations: an {I}ntroduction,
  2010.
\newblock Available on semanticscholar.org.

\bibitem[WCS{\etalchar{+}}18]{wang2018deep}
Haojie Wang, Han Chen, Agus Sudjianto, Richard Liu, and Qi~Shen.
\newblock Deep learning-based {B}{S}{D}{E} solver for {L}{I}{B}{O}{R} market
  model with application to bermudan swaption pricing and hedging.
\newblock {\em arXiv preprint arXiv:1807.06622}, 2018.
\newblock Also available at SSRN: https://ssrn.com/abstract=3214596 or
  http://dx.doi.org/10.2139/ssrn.3214596.

\bibitem[YXS19]{yu2019deep}
Bing Yu, Xiaojing Xing, and Agus Sudjianto.
\newblock Deep-learning based numerical {B}{S}{D}{E} method for {B}arrier
  options.
\newblock {\em arXiv preprint arXiv:1904.05921}, 2019.
\newblock Also available at SSRN: https://ssrn.com/abstract=3366314 or
  http://dx.doi.org/10.2139/ssrn.3366314.

\end{thebibliography}

\end{document}